\documentclass[aps,prb,floatfix,twocolumn,amsmath,amssymb,showpacs]{revtex4-1}

\usepackage{graphicx}
\usepackage{bm}
\usepackage{color}
\usepackage{epstopdf}
\usepackage{array}
\usepackage{mathtools}
\usepackage{verbatim}

\begin{document}
\title{Phonon hydrodynamics, thermal conductivity and second sound in 2D crystals}

\author{P. Scuracchio}
 \email{pablo.scuracchio@uantwerpen.be}
\author{K. H. Michel}
 \email{ktdm@skynet.be}
\author{F. M. Peeters}
 \email{francois.peeters@uantwerpen.be}
\affiliation{
Universiteit Antwerpen, Department of Physics, Groenenborgerlaan 171, Be-2020 Antwerpen, Belgium}

\date{\today}


\begin{abstract}

Starting from our previous work where we have obtained a system of coupled integro-differential equations for acoustic sound waves and phonon density fluctuations in 2D crystals, we derive here the corresponding hydrodynamic equations and study their consequences as function of temperature and frequency. These phenomena encompass propagation and damping of acoustic sound waves, diffusive heat conduction, second sound and Poiseuille heat flow, all of which are characterized by specific transport coefficients. We calculate these coefficients by means of correlation functions without using the concept of relaxation time. Numerical calculations are performed as well in order to show the temperature dependence of the transport coefficients and of the thermal conductivity. As a consequence of thermal tension mechanical and thermal phenomena are coupled. We calculate the dynamic susceptibilities for displacement and temperature fluctuations and study their resonances. Due to the thermo-mechanical coupling the thermal resonances such as Landau-Placzek peak and second sound doublet appear in the displacement susceptibility and conversely the acoustic sound wave doublet appears in the temperature susceptibility, Our analytical results do not only apply to graphene but are also valid for arbitrary 2D crystals with hexagonal symmetry like 2D h-BN, 2H-transition metal dichalcogenides and oxides.   
\end{abstract}

\maketitle

%
\section{Introduction}
\label{sec:introduction}
%

Phonon hydrodynamic phenomena such as propagation and attenuation of sound waves\cite{landau}, thermal conductivity\cite{peierlsfirst,peierls,casimir,klemens,leibfriedscho,ziman}, Poiseuille flow\cite{sussmann,gurzhi,mezhov, guyer1} and second sound\cite{ward1,ward2,dingle,sussmann,prohofsky,mezhov,ackerman,guyer,GM1,GM2,guyer1} in insulating 3D crystals have been investigated by experiment and theory since a long time. From a microscopic point of view these phenomena are closely related to anharmonic phonon interaction\cite{peierls,ziman}. Recently the subject has known a revival due to the discovery of 2D crystals\cite{novo}. Layered and 2D crystals exhibit a peculiar anharmonic interaction, linear in in-plane displacements and quadratic in out-of-plane displacements (flexural modes).This leads to unusual static effects such as negative thermal expansion\cite{liftshitz,mounet} and is also at the origin of unusual high thermal conductivity in graphene\cite{balandin1,seol}. In particular, it was found\cite{seol,lindsay} that the ZA (flexural acoustic) modes carry most of the heat in suspended graphene and that the selection rule for anharmonic phonon scattering in 2D crystals restricts momentum dissipating umklapp scattering. Numerical calculations using the single-mode relaxation time approximation confirm that thermal transport is dominated by ZA modes\cite{bonini}.

A study of phonon lifetimes as function of the wave-vector throughout the 2D Brillouin zone has shown that in a broad temperature range below room temperature the decay rate of flexural modes is much less affected by umklapp processes than the decay rate of in-plane modes\cite{seba}. The distinctive role played by flexural modes in anharmonic scattering processes has stimulated recent theoretical studies of hydrodynamic phenomena such as second sound and Poiseuille flow in 2D crystals\cite{lee_naturecom,cepe_naturecom,cepe_prx}. In these works the derivation of hydrodynamic phenomena is based on the solution of a linearized  Peierls-Boltzmann transport equation (kinetic equation) for the space and time dependent phonon density distribution, either by  iterative methods\cite{lee_naturecom} or by a variational approach\cite{cepe_naturecom,cepe_prx}. It was concluded that in 2D crystals hydrodynamic phenomena should exist in a broad temperature range up to and above room temperature. For a most recent review of the literature and discussion on thermal transport in 2D materials, see Ref. (\onlinecite{revchinos}).

While in the aforementioned studies all emphasis has been concentrated on the kinetic equation, the present authors have recently derived a system of dynamic equations for in-plane displacement correlations coupled to phonon density fluctuations\cite{SMP}. A linear coupling between in-plane lattice displacements and ZA phonon density fluctuations is a direct consequence of the anharmonic interaction mentioned before. As a result of linear response theory and Green's functions techniques we find that phonon density fluctuations enter as an additional driving term in the sound wave equation for in-plane lattice deformations while the kinetic equation contains an additional term due to coherent lattice displacements. This calls for a study of the ensuing thermo-mechanical effects on phonon hydrodynamics in 2D crystals. Such an investigation paves also the way for a broader use of experimental techniques\cite{griffin} since thermal effects will appear in mechanical response and vice-versa.

The content of this paper is the following. In Sec. \ref{sec:basicconcepts} we recall the basic concepts and results of our previous work\cite{SMP}. In Sec. \ref{sec:solutions} we present solutions for the coupled system of dynamic equations, first by assuming only energy conservation in the collision term of the kinetic equation, and second by taking also into account approximate conservation of crystal momentum. In the former case temperature variations will be described by a diffusion equation coupled to the in-plane lattice deformations, in the latter case depending on the frequency of the perturbation, temperature fluctuations will be described either by a wave equation (second sound) coupled to lattice deformations (first sound) or by Poiseuille flow. In both cases the equation for in-plane sound waves will be coupled to temperature fluctuations. In the steady-state case phonon transport in a confined crystal is described by Poiseuille flow. In Sec. \ref{sec:respfunc} we derive thermal and mechanical dynamic response functions and study their resonances in the diffusive and second sound regime. In Sec. \ref{sec:transpcoe} we calculate transport coefficients such as lattice viscosity, kinematic phonon viscosity and thermal diffusion by using a correlation function method developed by G\"otze and one of the present authors\cite{GM3}. In Sec. \ref{sec:numresults} we present numerical evaluations of these transport coefficients as functions of temperature for the case of graphene. Concluding remarks in Sec. \ref{sec:conclusions} and a comment on recent experiments close the paper.  

%
\section{Basic concepts}
\label{sec:basicconcepts}
%

We recall some basic concepts and results from our previous work\cite{SMP} on sound waves and flexural mode dynamics in 2D hexagonal crystals. The previous paper is hereafter referred to as I. The Hamiltonian operator which describes the coupling between acoustic in-plane and out-of-plane displacements was defined as $H=H_h + \Phi^{(3)}$. In terms of normal coordinates $Q^{\alpha}_{\vec{q}}$ and their conjugate momenta $P^{\alpha}_{\vec{q}}$ one has
\begin{align}
     &H_h = \frac{1}{2} \sum_{\vec{q},\alpha} \Big[ {P^{\alpha}_{\vec{q}}}^\dagger P^{\alpha}_{\vec{q}} + \omega^2 (\vec{q},\alpha)  {Q^{\alpha}_{\vec{q}}}^\dagger Q^{\alpha}_{\vec{q}} \Big] \; , \notag \\
     &\Phi^{(3)} = \frac{1}{2} \sum_{\vec{p},\vec{k},\vec{q},i} \Phi^{(3)} \left(\! \begin{array}{ccc} i & \zeta & \zeta \\ \vec{q} & \vec{k} & \vec{p} \end{array} \! \right) Q^{i}_{\vec{q}} Q^{\zeta}_{\vec{k}} Q^{\zeta}_{\vec{p}} \; .
\label{eqn:hamiltonians}
\end{align}
The wave vectors $\vec{q}$, $\vec{k}$, and $\vec{p}$ all belong to the first 2D Brillouin zone (BZ) and $\alpha = \{ 1,2, \zeta \}$ refers to the 3 acoustic polarizations LA (longitudinal acoustic in-plane), TA (transverse acoustic in-plane) and ZA (flexural acoustic), respectively. It is understood that $\omega^2 (\vec{q},\alpha)$ is the renormalized flexural phonon frequency and hence $\omega (\vec{q},\zeta)$ is linear in $\vec{q}$ at long wavelengths\cite{seba}. Along this paper, polarizations denoted with Latin indexes $i, j = \{ 1,2 \}$ refer only to in-plane modes. 

At long wavelengths, the Fourier-transformed center of mass displacements per unit cell $s_i(\vec{q})$ are related to the normal coordinates by
\begin{align}
    s_i(\vec{q}) = \frac{1}{\sqrt{m}} Q^{\alpha}_{\vec{q}}  \delta_{\alpha i} \; ,
   \label{eqn:centerofmass}
\end{align}
where $m$ is the total mass per unit cell, i. e. $m=2m_C$, $m_B + m_N$ or $m_{Mo} + 2 m_S$ in the case of graphene, h-BN or MoS$_2$. The corresponding conjugate momenta are
\begin{align}
    p_i(\vec{q}) = \sqrt{m} P^{\alpha}_{\vec{q}}  \delta_{\alpha i} \; .
   \label{eqn:centerofmass2}
\end{align}
The third order anharmonic coupling, calculated for the case of a central force interaction potential between nearest neighbors in a 2D hexagonal crystal of $N$ unit cells is given by
\begin{align}
    &\Phi^{(3)} \left( \! \begin{array}{ccc} i & \zeta & \zeta \\ \vec{q} & \vec{k} & \vec{p} \end{array} \! \right) = \frac{i8}{\sqrt{Nm^3}} \sum_s \phi^{(3)}_{i z z} (A_1 ; B_s) \notag\\
    &\times \text{cos} \Big[ \frac{(\vec{q}+\vec{k}+\vec{p}) \cdot \vec{\rho}(B_s)}{2} \Big] \text{sin} \Big[ \frac{\vec{q} \cdot \vec{r}(B_s)}{2} \Big] \text{sin} \Big[ \frac{\vec{k} \cdot \vec{r}(B_s)}{2} \Big] \notag\\
    &\times \text{sin} \Big[ \frac{\vec{p} \cdot \vec{r}(B_s)}{2} \Big] \Delta (\vec{q}+\vec{k}+\vec{p}) \; .
  \label{eqn:anharmonichamiltonian}
\end{align}
Here the interaction parameters $\phi^{(3)}_{i z z} (A_1 ; B_s)$ and the vectors $\vec{r}(B_s)$, $\vec{\rho}(B_s)$ have been defined in I. The crystal momentum conservation function  $\Delta(\vec{q}+\vec{k}+\vec{p}) = \sum_{\vec{G}} \: \delta_{\vec{q}+\vec{k}+\vec{p},\vec{G}}$ accounts for normal (N-processes) and umklapp processes (U-processes) when the reciprocal 2D lattice vector $\vec{G}=0$ or $\vec{G} \neq 0$, respectively.

In terms of phonon creation and annihilation operators $b^{\alpha \dagger}_{\vec{q}}$ and $b^\alpha_{\vec{q}}$, the normal coordinates are written as
\begin{align}
     Q^{\alpha}_{\vec{q}} &=\sqrt{\frac{\hbar}{2\omega(\vec{q},\alpha)}}(b^{\alpha \dagger}_{-\vec{q}}+b^{\alpha}_{\vec{q}}) \; , \notag\\
     P^{\alpha}_{\vec{q}} &=-i \sqrt{\frac{\hbar \omega(\vec{q},\alpha)}{2}}(b^{\alpha}_{\vec{q}} - b^{\alpha \dagger}_{-\vec{q}}) \; ,
  \label{eqn:normaloperatosr}
\end{align}
where $b^{\alpha}_{\vec{q}}$ and $b^{\alpha \dagger}_{\vec{q}}$ satisfy the usual commutation rules for Bose operators. Using linear response theory and thermal Green's functions techniques we have derived in I the following system of coupled dynamic equations. The first one describes sound waves due to coherent in-plane lattice deformations $s_i (\vec{q}, \omega)$:
\begin{align}
    &\Big[ m \omega^2 \delta_{ij}  - v_{2D} q_k q_l C_{ik,jl} \Big] s_j(\vec{q},\omega) = -F_i(\vec{q},\omega) \notag \\ 
    &+ i q_j \frac{\hbar}{N} \sum_{\vec{k},\alpha} \delta_{\alpha \zeta} h_{ji} (\vec{k},\zeta) n^\prime (\vec{k},\alpha) \nu(\vec{k},\alpha ; \vec{q},\omega)  \; .
\label{eqn:soundwaves}
\end{align}
The second one is a kinetic equation
\begin{align}
        &\Big[ \omega-q_i v_i(\vec{k},\alpha) \Big] \nu(\vec{k},\alpha;\vec{q},\omega) + i \hbar \omega q_i \delta_{\alpha \zeta} h_{ij}(\vec{k},\zeta) s_j(\vec{q},\omega) \notag\\
    &+ q_i v_i(\vec{k},\alpha) \hbar \omega(\vec{k},\alpha) \frac{\Theta(\vec{q},\omega)}{T} =  
    -i \hat{C} \nu(\vec{k},\alpha;\vec{q},\omega) \; ,
\label{eqn:nuflex}
\end{align}
where the quantity $\nu(\vec{k}, \alpha ; \vec{q}, \omega)$ describes the deviation from local equilibrium of the dynamic phonon density distribution (see Eq. (\ref{eqn:expansiondensity}) below). Both equations are coupled by the thermoelastic coupling
\begin{align}
    h_{ij} (\vec{k},\zeta) &=  \frac{1}{2 m \omega_R (\vec{k},\zeta)} \sum_s \phi^{(3)}_{izz} (A_1 ; B_s) \notag \\ 
    &\times  (\vec{k} \cdot \vec{r}(B_s))^2 r_j (B_s) \; .
    \label{eqn:definitioncompleteofhij}
\end{align}
In agreement with Onsager's reciprocity principle the coupling is symmetric in Eqs. (\ref{eqn:soundwaves}) and (\ref{eqn:nuflex}). The structure of the anharmonic coupling $\Phi^{(3)}$, Eq. (\ref{eqn:hamiltonians}), implies that only the flexural phonon density deviation $n^{\prime}(\vec{k}, \zeta) \nu(\vec{k}, \zeta; \vec{q}, \omega)$ is coupled to the in-plane deformations. The factor $n^{\prime}(\vec{k}, \zeta)$ stands for
\begin{align}
    n^\prime (\vec{k},\alpha) = \frac{n(\vec{k},\alpha)(1+n(\vec{k},\alpha))}{k_B T} = -\frac{1}{\hbar} \frac{\partial n(\vec{k},\alpha)}{\partial \omega (\vec{k},\alpha)} \; ,
\label{eqn:nprime}
\end{align}
where $k_B$ is the Boltzmann constant and $n(\vec{k},\alpha) = [\textrm{exp}(\hbar \omega / k_B T) - 1]^{-1}$ is the equilibrium Bose-Einstein distribution for acoustic phonons of energy $\hbar \omega (\vec{k}, \alpha)$, at equilibrium temperature $T$. The remaining quantities occurring in Eqs. (\ref{eqn:soundwaves}) and (\ref{eqn:nuflex}) are the Fourier transformed external mechanical force $F_i (\vec{q}, \omega)$ and temperature perturbation $\Theta (\vec{q}, \omega)$, the collision operator $\hat{C}$ (see Eqs. (\ref{eqn:coldefgen}) - (\ref{eqn:coldef2}) below), the isothermal elastic constants $C_{ik,jl}$, the phonon group velocity $v_i (\vec{k}, \alpha) = \partial \omega (\vec{k}, \alpha) / \partial k_i$ and the area of the crystal unit cell $v_{2D}$. As shown in I, $\nu (\vec{k}, \alpha ; \vec{q}, \omega)$ is related to the non-equilibrium phonon density $n (\vec{k}, \alpha ; \vec{q}, \omega)$ by 
\begin{align}
n(\vec{k}, \alpha ; \vec{q},\omega) = \tilde{n}(\vec{k}, \alpha ; \vec{q},\omega) + n^\prime(\vec{k}, \alpha) \nu(\vec{k}, \alpha ; \vec{q},\omega) \; ,
\label{eqn:expansiondensity}
\end{align}
where
\begin{align}
\tilde{n}(\vec{k}, \alpha ; \vec{q},\omega) = (\textrm{exp}(\hbar \omega(\vec{k}, \alpha ; \vec{q},\omega) / k_B T) - 1)^{-1} \; ,
\end{align}
is the local equilibrium phonon distribution with frequency
\begin{align}
\omega(\vec{k}, \alpha ; \vec{q},\omega) = \omega(\vec{k}, \alpha) - i \delta_{\alpha \zeta} q_i \frac{h_{ij} (\vec{k},\zeta)}{\sqrt{N}} s_j (\vec{q}, \omega)  \; .
\label{eqn:omegaflashera}
\end{align}
As originally pointed out by Akhiezer\cite{akhiezer}, a sound wave passing through a crystal causes a disturbance of the distribution of thermal phonons\cite{bomel,woodruff,GM1}. It was shown in I that as a consequence of collisions (quadratic effects in $\Phi^{(3)}$) the in-plane phonon density is driven out of equilibrium too. The collision term in the kinetic equation (\ref{eqn:nuflex}) then stands for 
\begin{align}
    \hat{C} \nu(\vec{k},\alpha;\vec{q},\omega) &= C^{(1)} \nu(\vec{k},\alpha;\vec{q},\omega) \delta_{\alpha, \zeta} \notag  \\
    &+ (1-\delta_{\alpha \zeta}) C^{(2)} \nu(\vec{k},\alpha;\vec{q},\omega) \; ,
\label{eqn:coldefgen}
\end{align}
where
\begin{align}
    &C^{(1)} \nu(\vec{k},\zeta)= \frac{1}{n^{\prime}(\vec{k}, \zeta)} \times \notag\\
    &\sum_{\vec{p} \vec{h},i} \bigg\{ W \left( \! \begin{array}{ccc} \zeta & \zeta & i \\ \vec{h} & \vec{k} & \vec{p} \end{array} \! \right) \Big[ \nu(\vec{k},\zeta) - \nu(\vec{h},\zeta) + \nu(\vec{p},i)\Big] \nonumber \\ 
    &+W \left( \! \begin{array}{ccc} i & \zeta & \zeta \\ \vec{p} & \vec{h} & \vec{k} \end{array}\! \right) \Big[ \nu(\vec{k},\zeta) - \nu(\vec{p},i) + \nu(\vec{h},\zeta)\Big] \nonumber \\
    &+W \left( \! \begin{array}{ccc} \zeta & i & \zeta \\ \vec{k} & \vec{p} & \vec{h} \end{array} \! \right) \Big[ \nu(\vec{k},\zeta) - \nu(\vec{p},i) - \nu(\vec{h},\zeta)\Big] \bigg\} \; ,
\label{eqn:coldef1}
\end{align}
and
\begin{align}
    &C^{(2)} \nu(\vec{k},j) = \frac{1}{n^{\prime}(\vec{k}, j)} \times \notag\\
    &\sum_{\vec{p} \vec{h}}  \bigg\{ \frac{1}{2} W \left( \! \begin{array}{ccc} j & \zeta & \zeta \\ \vec{k} & \vec{p} & \vec{h} \end{array} \! \right) \big[ \nu(\vec{k},j) - \nu(\vec{p},\zeta) - \nu(\vec{h},\zeta)\big] \nonumber \\ 
    &+W \left( \! \begin{array}{ccc} \zeta & j & \zeta \\ \vec{p} & \vec{k} & \vec{h} \end{array} \! \right) \Big[ \nu(\vec{k},j) - \nu(\vec{p},\zeta) + \nu(\vec{h},\zeta)\Big] \bigg\} \; .
\label{eqn:coldef2}
\end{align}
Here and in the following we simplify the actual notation by not mentioning explicitly the $\vec{q}, \omega$ dependence of the quantity $\nu(\vec{k},\alpha)$. We recall that expressions (\ref{eqn:coldefgen})- (\ref{eqn:coldef2}) correspond to linearized collision terms of Peierls-Boltzmann type kinetic equations. The transition probabilities $W$ in (\ref{eqn:coldef1}) and (\ref{eqn:coldef2}) are given by
\begin{align}
    W \left( \! \begin{array}{ccc} \zeta & j & \zeta \\ \vec{k} & \vec{p} & \vec{h} \end{array} \! \right)&= 2 \pi \hbar \; \bigg\vert \Psi^{(3)} \left(\! \begin{array}{ccc} \zeta & j & \zeta \\ -\vec{k} & \vec{p} & \vec{h} \end{array} \! \right) \bigg\vert^2 \notag\\
    &\times \sqrt{k_B T n^{\prime}(\vec{k},\zeta)n^{\prime}(\vec{p},j)n^{\prime}(\vec{h},\zeta)} \notag\\
    &\times \delta(\omega(\vec{h},\zeta)-\omega(\vec{k},\zeta)+\omega(\vec{p},j)) \; , \notag\\
    W \left( \! \begin{array}{ccc} j & \zeta & \zeta \\ \vec{p} & \vec{h} & \vec{k} \end{array} \! \right)&= 2 \pi \hbar \; \bigg\vert \Psi^{(3)} \left(\! \begin{array}{ccc} j & \zeta & \zeta \\ -\vec{p} & \vec{h} & \vec{k} \end{array} \! \right) \bigg\vert^2 \notag\\
    &\times \sqrt{k_B T n^{\prime}(\vec{p},j)n^{\prime}(\vec{h},\zeta)n^{\prime}(\vec{k},\zeta)} \notag\\
    &\times \delta(\omega(\vec{h},\zeta)-\omega(\vec{p},j)+\omega(\vec{k},\zeta)) \; ,
\label{eqn:DefinitionW2}
\end{align}
with
\begin{align}
    \Psi^{(3)} \left(\! \begin{array}{ccc} j & \zeta & \zeta \\ -\vec{p} & \vec{h} & \vec{k} \end{array} \! \right)= \frac{\Phi^{(3)} \left(\! \begin{array}{ccc} j & \zeta & \zeta \\ -\vec{p} & \vec{h} & \vec{k} \end{array} \! \right)}{\sqrt{8 \omega(\vec{k},\zeta) \omega(\vec{p},j) \omega(\vec{h},\zeta)}} \; . 
\label{eqn:defpsimayus}
\end{align}
While the $\delta$-functions account for phonon energy conservation in scattering processes involving three phonons, the $\Delta$-function contained in $\Phi^{(3)}$ ensures that crystal momentum is only conserved up to a reciprocal lattice vector. Since U-processes involve large wave vectors close to the BZ's boundaries, the corresponding phonon energies $\hbar \omega_{BZ}$ are relatively large (see e.g. acoustic phonon dispersions in graphene)\cite{mohr} . The phonon occupation factors $n^{\prime}$ contained in the transition probabilities $W$ are only significant for temperatures $k_B T \approx \hbar \omega_{BZ}$. As it is well known (see e.g. Ziman\cite{ziman}), U-processes are frozen out as $\approx \textrm{e}^{-\Theta_D /T}$, with $\Theta_D$ of the order of the Debye temperature. With decreasing $T$ only N-processes survive. We recall that the hydrodynamic regime corresponds to a situation where the system is close to full thermodynamic equilibrium. Such a local equilibrium situation is established by N-processes while the subsequent decay to total equilibrium is described by U-processes or other crystal momentum destroying processes such as impurity and boundary scattering. 

While Eq. (\ref{eqn:soundwaves}) describes long wavelength and low frequency sound waves, the reduction of the kinetic equation (\ref{eqn:nuflex}) to the hydrodynamic case, i.e. to an equation for local temperature fluctuations requires an additional procedure. The principle is based on the existence of conservation laws and the solution of the kinetic equation in terms of the corresponding local densities\cite{Huang}. We observe that Eq. (\ref{eqn:nuflex}) is an inhomogeneous linear integral equation with an Hermitian kernel. Indeed defining a scalar product for any two functions $a_1(\vec{k},\alpha)$ and $a_2(\vec{k},\alpha)$ of the phonon variables as
\begin{align}
\langle a_1 \vert a_2 \rangle=\frac{1}{N} \sum_{\vec{p},\alpha} a_1^*(\vec{p},\alpha) n^\prime(\vec{p},\alpha) a_2(\vec{p},\alpha) \; ,
\label{eqn:scalarproduct}
\end{align}
we obtain for the collision term
\begin{align}
\langle \nu_1 \vert \hat{C} \vert \nu_2 \rangle = \langle \nu_2 \vert \hat{C} \vert \nu_1 \rangle \; ,
\label{eqn:symmetrycollision}
\end{align}
where 
\begin{align}
&\langle \nu_1 \vert \hat{C} \vert \nu_2 \rangle = \frac{1}{N} \sum_{\vec{p}} \Big\{ \nu_1^*(\vec{p},\zeta) C^{(1)} \nu_2(\vec{p},\zeta)  \notag\\
&+  \sum_j \nu_1(\vec{p},j)^* C^{(2)} \nu_2(\vec{p},j)  \Big\} \notag\\
&=\frac{1}{N} \sum_{\vec{k}\vec{p}\vec{h},j} \bigg\{ W \left( \! \begin{array}{ccc} \zeta & j & \zeta \\ \vec{k} & \vec{p} & \vec{h} \end{array} \! \right) \big[ \nu_1^*(\vec{k},\zeta) - \nu_1^*(\vec{h},\zeta) - \nu_1^*(\vec{p},j) \big]  \notag\\
&\times \big[ \nu_2(\vec{k},\zeta) - \nu_2(\vec{h},\zeta) - \nu_2(\vec{p},j) \big] \notag\\
&+\frac{1}{2} W \left( \! \begin{array}{ccc} j & \zeta & \zeta \\ \vec{p} & \vec{h} & \vec{k} \end{array} \! \right)  \big[ \nu_1^*(\vec{k},\zeta) + \nu_1^*(\vec{h},\zeta) - \nu_1^*(\vec{p},j)\big] \notag\\
&\times \big[ \nu_2(\vec{k},\zeta) + \nu_2(\vec{h},\zeta) - \nu_2(\vec{p},j)\big]
\bigg\} \; .
\label{eqn:collisionextectedvalue}
\end{align}
Consequently $\hat{C}$ is an Hermitian operator and admits a spectral representation
\begin{align}
\hat{C}=\sum_l \omega^l \frac{\vert \chi^l \rangle \langle \chi^l \vert}{\langle \chi^l \vert \chi^l \rangle}  \; ,
\label{eqn:espectralcollision}
\end{align}
where $\omega^l$ and $\chi^l$ are the corresponding eigenvalues and orthogonalized eigenfunctions. Since the transition probabilities satisfy $W \geq 0$, the eigenvalues $\omega^l$ (also called relaxation frequencies) satisfy $\omega^l \geq 0$. From the conservation of energy in the phonon scattering  processes Eq. (\ref{eqn:DefinitionW2}) it follows that
\begin{align}
    \chi^0(\vec{k},\alpha) = \frac{\hbar}{T} \omega (\vec{k}, \alpha) \; ,
\label{eqn:defchizero}
\end{align}
is an eigenfunction of $\hat{C}$ with eigenvalue $\omega^0 = 0$. Strictly speaking there are no further conservation laws for the scattering processes. However at low temperature where U-processes are frozen out, crystal momentum $\hbar \vec{k}$ can be treated as a conserved quantity. 

%
\section{Hydrodynamic equations}
\label{sec:solutions} 
%

The solution  of an inhomogeneous linear integral equation with an Hermitian kernel can be obtained by means of a series expansion in terms of the eigenfunctions\cite{courant}. We will start from an expansion of the non equilibrium phonon density $\nu(\vec{k}, \alpha; \vec{q},\omega)$ in terms of the eigenfunctions $\chi^l (\vec{k},\alpha)$ of the collision operator. We proceed in two steps, first (subsection A) we will assume that only phonon energy is conserved in collisions and second (subsection B) we will in addition treat crystal momentum as an almost conserved quantity. 

%
\subsection{Energy Conservation}
\label{subsec:energyconservation}
%

The non equilibrium phonon density is written as
\begin{align}
    \nu(\vec{k},\alpha;\vec{q},\omega)= \theta (\vec{q},\omega) \chi^0(\vec{k},\alpha)+ \sum_{l > 0} a^l (\vec{q},\omega) \chi^l (\vec{k},\alpha) \; .
\label{eqn:expansioneigenvec}
\end{align}
Here we have separated off the coefficient $a^0 (\vec{q},\omega) = \theta (\vec{q},\omega)$ which refers to phonon energy conservation. In the following we will calculate the resolvent of the integral equation (\ref{eqn:nuflex}) by using perturbation theory. Following Ref. (\onlinecite{GM2}) we define the hydrodynamic regime as that one for which the external frequency $\omega$ (and corresponding wave vector $\vec{q}$) is small in comparison to the relaxation frequencies $\omega^l$. We consider the situation of high $T$ such that a distinction between N-processes and U-processes is irrelevant. Then $\chi^l (\vec{k},\omega)$ with $l > 0 $ exhausts the spectrum of crystal momentum conserving and destroying scattering processes. Recalling from Sec. \ref{sec:basicconcepts} that $\nu$ has the dimension of energy, we see that the quantity $\theta (\vec{q}, \omega)$ has the dimension of temperature. In order to perform a reduction from the $\vec{k}, \alpha$ dependent phonon density $\nu (\vec{k}, \alpha; \vec{q}, \omega)$ to an equation for the temperature fluctuations $\theta (\vec{q}, \omega)$, we insert expression (\ref{eqn:expansioneigenvec}) into Eqs. (\ref{eqn:soundwaves}) and (\ref{eqn:nuflex}). We then isolate the quantities $\theta (\vec{q}, \omega)$ and $a^l (\vec{q}, \omega)$ by multiplying Eq. (\ref{eqn:nuflex}) scalarly with $\chi^0 (\vec{k}, \alpha)$ and $\chi^l (\vec{k}, \alpha)$ respectively. In evaluating the scalar products we take into account the parity of the relevant functions: while $\omega (\vec{k}, \alpha)$ and $h_{ij} (\vec{k}, \zeta)$ are even in $\vec{k} \to - \vec{k}$, $v_i (\vec{k}, \alpha)$ is uneven. 

Defining the specific heat per unit cell as
\begin{align}
c_v = \frac{\hbar^2}{T} \langle \omega \vert \omega \rangle \; ,
\label{eqn:groundstatenorm}
\end{align}
and the thermal tension\cite{seba} by
\begin{align}
\beta_{ij} &= \frac{\hbar^2}{T v_{2D}} \langle h_{ij} \vert \omega \rangle \notag\\
&=  \frac{\hbar^2}{Nv_{2D}T} \sum_{\vec{k}} h_{ji} (\vec{k},\zeta) n^\prime (\vec{k},\zeta) \omega(\vec{k},\zeta) \; ,
\label{eqn:betadef}
\end{align}
we obtain from Eq. (\ref{eqn:soundwaves})
\begin{align}
    &\Big[m \omega^2 \delta_{ij}  - v_{2D} q_k q_l C_{ik,jl} \Big] s_j(\vec{q}, \omega) = - F_i (\vec{q}, \omega) \notag\\
    &+ i v_{2D} q_j \beta_{ji}  \theta(\vec{q},\omega) + i \hbar q_j \sum_{l > 0} \langle h_{ji} \vert \chi^l \rangle a^l (\vec{q},\omega) \; ,
\label{eqn:soundwaves2}
\end{align}
and from Eq. (\ref{eqn:nuflex})
\begin{align}
    \omega \theta(\vec{q},\omega) - q_i \hbar \sum_{l > 0} &\frac{\langle \omega v_i \vert \chi^l \rangle}{c_v} a^l (\vec{q},\omega) \notag\\
    &+ i \omega v_{2D} q_i \beta_{ij} s_j(\vec{q},\omega)\frac{T}{c_v} =0 \; ,
\label{eqn:nuflex0}
\end{align}
and
\begin{align}
    &(\omega+i\omega^l) \langle \chi^l \vert \chi^l \rangle a^l (\vec{q},\omega) +i \hbar \omega q_i \langle \chi^l \vert h_{ij}\rangle s_j(\vec{q},\omega) \notag\\
    &= q_i \hbar \frac{\langle \chi^l \vert v_i \omega \rangle}{T} [ \theta (\vec{q},\omega) - \Theta(\vec{q},\omega) ] \; .
  \label{eqn:nuflexmuarranged}
\end{align}
Taking into account that $0 \leq \omega < \omega^l$ we use the approximation
\begin{align}
    a^l (\vec{q},\omega) &=\frac{1}{\omega^l \langle \chi^l \vert \chi^l \rangle} \{- \hbar \omega q_i \langle \chi^l \vert h_{ij}\rangle s_j(\vec{q},\omega) \notag\\
    &- i q_i \hbar \frac{\langle \chi^l \vert v_i \omega \rangle}{T} [ \theta (\vec{q},\omega) - \Theta(\vec{q},\omega) ]\}  \; ,  
  \label{eqn:amudespeje1}
\end{align}
and eliminate $a^l (\vec{q}, \omega)$ in Eqs. (\ref{eqn:soundwaves2}) and (\ref{eqn:nuflex0}) with the following results:
\begin{align}
    &\Big[m \omega^2 \delta_{ij}  - v_{2D} q_k q_l (C_{ik,jl}-i \omega \eta_{ik,jl}) \Big] s_j(\vec{q},\omega) =  \notag\\
    &- F_i (\vec{q},\omega) + i v_{2D} q_j \beta_{ji}  \theta(\vec{q},\omega) \; ,
\label{eqn:soundwaves3}
\end{align}
and
\begin{align}
    (\omega+iq_i q_k \lambda_{ik})&\theta(\vec{q},\omega) = i q_i q_k \lambda_{ik} \Theta(\vec{q},\omega) \notag\\
    &- i \omega q_i \beta_{ij} v_{2D} T  c_v^{-1} s_j(\vec{q},\omega) \; .
\label{eqn:nuflexcomplete}
\end{align}
Here we have introduced the tensors
\begin{align}
    \eta_{ij,kl} &= \frac{\hbar^2}{v_{2D}}  \sum_{l>0} \frac{\langle h_{ij} \vert \chi^l \rangle \langle \chi^l \vert h_{kl} \rangle}{ \omega^l \langle \chi^l \vert \chi^l \rangle } \; ,  \label{eqn:viscositytensordef} \\
    \lambda_{ik} &= \hbar^2 \sum_{l >0} \frac{\langle \omega v_i \vert \chi^l \rangle \langle \chi^l \vert v_k \omega \rangle}{T c_v  \omega^l \langle \chi^l \vert \chi^l \rangle} \; . \label{eqn:lambdatensordef}
\end{align}
As already mentioned, $\theta(\vec{q},\omega)$ has the dimension of temperature. Taking the static limit ($\omega \to 0$) of Eq. (\ref{eqn:nuflexcomplete}), we see that $\theta(\vec{q},\omega=0)$ is equal to the external temperature source $\Theta(\vec{q},\omega=0)$. We therefore infer that $\theta(\vec{q},\omega)$ is the local temperature in the hydrodynamic regime. Eq. (\ref{eqn:nuflexcomplete}) has the meaning of a temperature diffusion equation where the quantity $\lambda_{ij}$ is the thermal diffusion coefficient. On the right hand side of this equation, beside the external temperature source, the lattice deformations $s_i (\vec{q}, \omega)$ act as an additional perturbation. The coupling between thermal and mechanical quantities, in casu local temperature and coherent lattice deformations respectively, is mediated by the thermal tension coefficient $\beta_{ij}$. Similarly, in the equation of motion (\ref{eqn:soundwaves3}) for the in-plane lattice deformations, in addition to the elastic restoring forces there appears now a dissipation term with dynamic viscosity $\eta_{ij,kl}$, while the local temperature variations, again mediated by the thermal tension, act as an additional driving force.

After inverse Fourier-transforming we rewrite Eqs. (\ref{eqn:soundwaves3}) and (\ref{eqn:nuflexcomplete}) as hydrodynamic equations where $s_i (\vec{r}, t)$ and $\theta (\vec{r}, t)$ are slowly space and time dependent variables describing the deformation of the crystalline lattice and the changes of the temperature respectively:
\begin{align}
    &m \partial^2_t s_i(\vec{r},t) - v_{2D} (C_{ik,jl} - \eta_{ik,jl} \partial_t)  \partial_k \partial_l s_j(\vec{r},t)   \notag\\
    &= - F_i (\vec{r},t) - v_{2D} \beta_{ji} \partial_j \theta (\vec{r},t) \; ,
\label{eqn:soundwaves3real}
\end{align}
and
\begin{align}
    \partial_t \theta(\vec{r},t) &=  \lambda_{ik} \partial_i \partial_k [ \theta(\vec{r},t)-\Theta(\vec{r},t) ] \notag\\
    &- \beta_{ij} v_{2D} T  c_v^{-1} \partial_t \partial_i s_j(\vec{r},t) \; .
\label{eqn:nuflexcompletereal}
\end{align}
Here $\partial_j$ stands for $\partial / \partial r_j$ and $\partial_t = \partial / \partial t$. From now on, this scenario will be addressed as the diffusion regime (DR).

In absence of an external heat source and of coupling to the lattice the last equation reduces to the Fourier temperature diffusion equation. In the static limit ($\partial_t = 0$) and in absence of mechanical forces, Eqs. (\ref{eqn:soundwaves3real}) and (\ref{eqn:nuflexcompletereal}) reduce to
\begin{align}
    \theta (\vec{r}, t) &= \Theta (\vec{r}, t) \; , \notag\\
    \partial_l  s_k(\vec{r},t) &=  C_{ij,kl}^{-1} \beta_{ij}  \Theta(\vec{r},t) \; ,
\label{eqn:soundwaves3staticlimitreal}
\end{align}
i.e. the equation of thermal expansion. We observe that so far we did not make use of hexagonal symmetry, hence Eqs. (\ref{eqn:soundwaves3}) and (\ref{eqn:nuflexcomplete}) or equivalently (\ref{eqn:soundwaves3real}) and (\ref{eqn:nuflexcompletereal}) are valid for 2D crystals of any symmetry.

%
\subsection{Energy and crystal momentum conservation}
\label{subsec:energyandmomentumconservation}
%

In subsection \ref{subsec:energyconservation} we have shown that if only phonon energy is conserved in scattering processes, temperature changes satisfy a diffusion equation. Here we will assume a situation 
where U-processes and other crystal momentum destroying processes such as phonon scattering with static impurities are relatively unimportant, while N-processes are still efficient to establish local equilibrium. Treating crystal momentum as an almost conserved quantity we replace Eq. (\ref{eqn:expansioneigenvec}) by
\begin{align}
    \nu(\vec{k},\alpha;\vec{q},\omega)&= \theta (\vec{q},\omega) \chi^0(\vec{k},\alpha) + \sum_{i=1,2} a^i (\vec{q},\omega) \chi^i (\vec{k}) \notag\\
    &+ \sum_{l > 2} a^l (\vec{q},\omega) \chi^l (\vec{k},\alpha) \; , 
\label{eqn:expansioneigenvecalsomomentum}
\end{align}
where the crystal momentum 
\begin{align}
\chi^i (\vec{k}) = \hbar k_i
\label{eqn:defchimomentum}
\end{align}
is treated as an eigenfunction with eigenfrequency $\omega^i$ and where $a^i (\vec{q}, \omega)$ has the dimension of velocity. We assume that $ \chi^i (\vec{k})$ with $i=1,2$ exhausts the spectrum of momentum destroying processes and that $\chi^l (\vec{k},\alpha)$ with $l > 2 $ refers to N-processes. Such a separation is meaningful at low $T$ where U-processes freeze out. If in addition other momentum destroying processes such as elastic impurity scattering are included in the collision operator, we assume that the concentration of impurities is sufficiently weak such that the combined relaxation frequency $\omega^i$ still satisfies $\omega^i < \omega^l$, for $i=1,2$ and $l>2$. The meaning of the first two terms on the right hand side of Eq. (\ref{eqn:expansioneigenvecalsomomentum}) can be traced back by linearizing the local displaced or drifting phonon distribution:
\begin{align}
    n_L(\vec{k},\alpha;\vec{q},\omega)&= \big[ \textrm{exp} \big[\frac{\hbar \omega (\vec{k}, \alpha ; \vec{q}, \omega) - \hbar \vec{k} \cdot \vec{a} (\vec{q}, \omega)}{k_B (T + \theta (\vec{q}, \omega))} \big] -1 \big]^{-1} \; ,
\label{eqn:displaced}
\end{align}
where $\vec{a} (\vec{q}, \omega)$ is the drift velocity, and $\omega(\vec{k},\alpha;\vec{q},\omega)$, Eq. (\ref{eqn:omegaflashera}), is the acoustic phonon frequency in a deforming lattice. The last term in Eq. (\ref{eqn:expansioneigenvecalsomomentum}) accounts for N-processes with relaxation frequencies $\omega^l$. It has been noticed that Eq. (\ref{eqn:displaced}) corresponds to a distribution function towards which normal processes relax\cite{woodruff} in a deformed lattice. In the steady-state case we recover the displaced phonon distribution with a constant drift velocity\cite{klemens,callaway}. 

In extending the procedure of subsection \ref{subsec:energyconservation} we now calculate $\theta$, $a^l$ and $a^i$. By taking the scalar products we use the fact that $\chi^i (\vec{k})$ is an uneven function of $\vec{k}$. Substituting $\nu (\vec{k}, \alpha ; \vec{q}, \omega)$, Eq. (\ref{eqn:expansioneigenvecalsomomentum}), into Eq. (\ref{eqn:nuflex}), multiplying scalarly by $\chi^0 (\vec{k}, \alpha)$ and eliminating $a^l$ by using now $0 \leq \omega < \omega^l$ with $l \geq 2$ we obtain
\begin{align}
    &(\omega + i q_i q_k \lambda^\prime_{ik} )\theta(\vec{q},\omega) = i q_i q_k \lambda^\prime_{ik} \Theta(\vec{q},\omega) \notag\\
    &- i \omega q_i \frac{T}{c_v} \beta_{ij} v_{2D} s_j (\vec{q},\omega)  + q_i d_{ik} a^k(\vec{q},\omega) \; .
\label{eqn:nu0flexcomplete2}
\end{align}
Here the quantities $\lambda^\prime_{ik}$ are defined by an expression similar to Eq. (\ref{eqn:lambdatensordef}) except that the $l$-sum does not include terms with $l=0,1, 2$ that refer to conserved quantities. In the last term on the right hand side we have defined
\begin{align}
d_{ij} = \frac{\langle \chi^0  \vert v_i \chi^j \rangle}{\langle \chi^0  \vert \chi^0 \rangle} =\frac{\hbar^2 }{c_v} \langle \omega  \vert v_i k_j \rangle \; .
\label{eqn:lambdaprimedef}
\end{align}
The quantities $a^i (\vec{q}, \omega)$ are found to satisfy
\begin{align}
    &\big[ (\omega + i \omega^i) \delta_{ik} + i q_j q_h \pi^\prime_{ij,kh} \big]  a^k(\vec{q},\omega) \notag\\ 
    &= q_j \frac{c_v d_{ji}}{T \langle \chi^i \vert \chi^i \rangle}  \big[ \theta(\vec{q},\omega) - \Theta(\vec{q},\omega) \big] \; ,
\label{eqn:nu0flexcomplete4}
\end{align}
for fixed $i$. Here we have defined the transport coefficient
\begin{align}
\pi^\prime_{ij,kh}=  \sum_{l > 2} \frac{\langle \chi^i v_j \vert \chi^l \rangle \langle \chi^l \vert v_h \chi^k \rangle  }{\langle \chi^l \vert \chi^l \rangle \omega^l \langle \chi^i \vert \chi^i \rangle} \; ,
\label{eqn:piprimedef}
\end{align}
which is due to N-processes and which plays the role of a kinematic viscosity of the phonon gas (see second sound regime and Poiseuille flow below). In writing down Eq. (\ref{eqn:nu0flexcomplete4}) we have neglected an additional transport coefficient which couples the phonon drift velocity to the lattice deformations $s_i$ since there is already a coupling to $s_i$ by the thermal tension in Eq. (\ref{eqn:nu0flexcomplete2}). Defining the matrix
\begin{align}
\Lambda_{ik} (\vec{q}, \omega) = (\omega + i \omega^i) \delta_{ik} + i q_j q_h \pi^\prime_{ij,kh} \; ,
\label{eqn:biglambdadef2}
\end{align}
we obtain
\begin{align}
a^k (\vec{q}, \omega)=  \Lambda^{-1}_{ki} (\vec{q}, \omega) \frac{c_v q_j d_{ji}}{ T \langle \chi^i \vert \chi^i \rangle} \big[  \theta (\vec{q}, \omega) - \Theta (\vec{q}, \omega) \big]  \; .
\label{eqn:ajdefdisplaced}
\end{align}
Though Eq. (\ref{eqn:ajdefdisplaced}) would be an appropriate tool to discuss the frequency dependence of the thermal conductivity\cite{majee} without resort to the Callaway model\cite{callaway}, we will not pursue this path here but restrict ourselves to the study of second sound and phonon Poiseuille flow.

In order to simplify the algebra, we will only retain the diagonal elements of the matrix $\Lambda_{ik}$ and restrict ourselves to a 2D crystal with hexagonal symmetry\cite{chaikin}. Then second rank tensors will reduce to scalars viz. $\lambda^{\prime}_{ij} = \lambda^{\prime} \delta_{ij}$,  $\beta_{ij} = \beta \delta_{ij}$,  $d_{ij} = d \delta_{ij}$, and $\omega^i \equiv \omega_V$, $i = \in \{ 1, 2 \}$, and Eq. (\ref{eqn:ajdefdisplaced}) becomes
\begin{align}
a^k (\vec{q}, \omega)=  \frac{c_v q_k d}{ T \langle \chi^k \vert \chi^k \rangle} \frac{ \big[  \theta (\vec{q}, \omega) - \Theta (\vec{q}, \omega) \big] }{(\omega + i \omega_V + i q_j^2  \pi^{\prime}_{kj,kj})} \; ,
\label{eqn:ajdefdisplaceddif}
\end{align}
which shows that the phonon drift velocity is damped by crystal momentum dissipating processes $\omega_V$ and by N-processes ($q^2 \pi^\prime $). In the following we will study the system of coupled equations (\ref{eqn:nu0flexcomplete2}) and (\ref{eqn:ajdefdisplaceddif}) for two distinct hydrodynamic frequency regimes: (i) $\omega_V < \omega < \omega^l$, and (ii) $\omega < \omega_V < \omega^l$.

\begin{figure}
\includegraphics[trim = 0.5cm 0cm 0cm 0.0cm, clip, width=8.7cm]{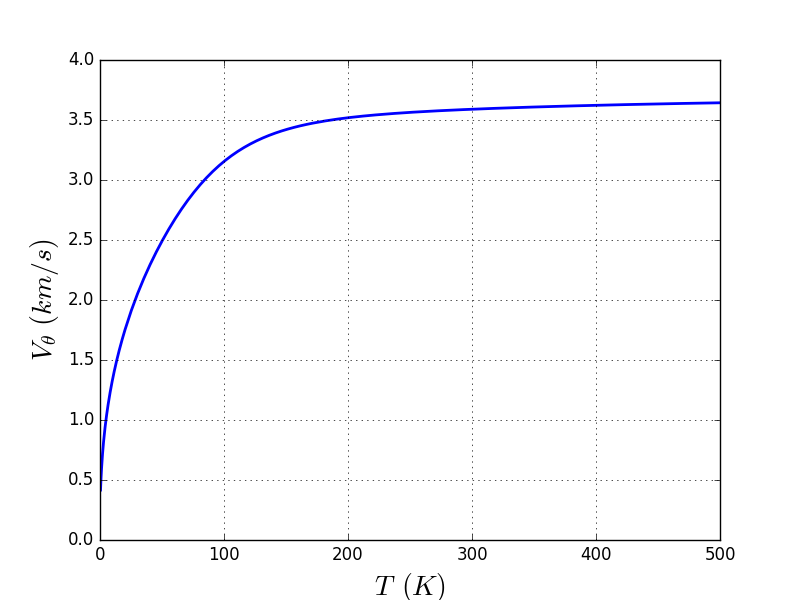}
\caption{Second sound velocity calculated for graphene according to Eq. (\ref{eqn:pdef}).}
\label{fig:ssvel}
\end{figure}

Case (i) corresponds to the frequency window condition\cite{GM2}, necessary for the existence of second sound. Substituting $a^k (\vec{q}, \omega)$ into Eq. (\ref{eqn:nu0flexcomplete2}), we get for a 2D crystal with hexagonal symmetry the wave equation
\begin{align}
    &\big[ \omega^2 - q^2 V^2_{\theta} + i \omega ( \omega_V + q^2 \lambda^\prime + q_j^2  \pi^\prime_{kj,kj}  )  \big] \theta(\vec{q},\omega) \notag\\
    &= \big[ - q^2 V^2_{\theta} + i \omega q^2 \lambda^\prime \big] \Theta(\vec{q},\omega)  \notag\\
    &- i \omega^2 \frac{T}{c_v} \beta v_{2D} q_i s_i (\vec{q},\omega) \; ,
\label{eqn:nu0flexcomplete5approx2secondsound}
\end{align}
where
\begin{align}
V^2_{\theta} = \frac{c_v}{T} \frac{d^2}{\langle \chi^k \vert \chi^k \rangle} \; ,
\label{eqn:pdef}
\end{align}
has the dimension of velocity squared and $q^2 = q_1^2 + q_2^2$. In obtaining this result we have neglected quadratic contributions in the dissipative terms such as $\omega_V \lambda^\prime$, $\pi^\prime \lambda^\prime$ as well as $\omega_V \beta$, $\pi^\prime \beta$. In analogy with the corresponding phenomenon originally studied in superfluid Helium\cite{khalatnikov}, a wave equation for temperature or equivalently phonon density fluctuations is called second sound. In Eq. (\ref{eqn:nu0flexcomplete5approx2secondsound}) the squared second sound velocity $V_\theta$ accounts for the restoring forces while the damping is determined by the relaxation frequency $\omega_V$ and the dissipation coefficients $\lambda^\prime$ and $\pi^\prime$ that are inversely proportional to the relaxation frequencies $\omega^l$ for N-processes. Besides the external temperature source, dynamic lattice deformations act as periodic temperature perturbations. The converse mechanism is readily studied by inserting the ansatz (\ref{eqn:expansioneigenvecalsomomentum}) into Eq. (\ref{eqn:soundwaves}) for the lattice displacements. Applying again the method of successive determination of $a^i$ and eliminating $a^l$ we obtain as result an equation similar to Eq. (\ref{eqn:soundwaves3}) where $\theta (\vec{q}, \omega)$ is determined by Eq. (\ref{eqn:nu0flexcomplete5approx2secondsound}). Transforming Eq. (\ref{eqn:nu0flexcomplete5approx2secondsound}) to a space and time dependent equation we get
\begin{align}
    &\big[ \partial_t^2 - V_{\theta}^2 \partial^2_j + [\omega_V -(\lambda^\prime + \pi^\prime_{kj,kj}) \partial^2_j ] \partial_t   \big] \theta (\vec{r},t) \notag\\
    &= \big[ - V_\theta^2  \partial^2_j - \lambda^\prime  \partial^2_j \partial_t \big] \Theta(\vec{r},t)  \notag\\
    &- \beta \frac{T}{c_v v_{2D}} \partial_t^2 \partial_i s_i (\vec{r},t) \; .
\label{eqn:nu0flexcomplete5approx2secondsoundreal}
\end{align}
The corresponding equation for the lattice deformation $s_i (\vec{r}, t)$ is of the same form as Eq. (\ref{eqn:soundwaves3real}) where now $\theta (\vec{r}, t)$ satisfies Eq. (\ref{eqn:nu0flexcomplete5approx2secondsoundreal}). This scenario will from now on be addressed as the second sound regime (SSR). In Fig. \ref{fig:ssvel} we plot the second sound velocity $V_\theta$ calculated for graphene as a function of temperature $T$. It ranges from $\sim 550$ ms$^{-1}$ at $T = 1$ K  to $\sim 3.6$ kms$^{-1}$ at $T = 430$ K and increases monotonically with temperature, in agreement with results of Lee et al.\cite{lee_naturecom}.

Turning to the case (ii) we take $\omega = 0$ and fix the external heat sources such that a constant temperature gradient is maintained in the crystal\cite{mezhov,gurzhi}. In real space Eq. (\ref{eqn:ajdefdisplaceddif}) leads to
\begin{align}
( \pi^{\prime}_{kj,kj} \partial^2_j - \omega_V ) a^k (\vec{r}) = \frac{c_v d}{T \langle \chi^k \vert \chi^k \rangle} \partial_k \theta (\vec{r}) \; .
\label{eqn:poisoille1}
\end{align}
This equation describes the steady-state flow of a phonon gas driven by a temperature gradient. In analogy with the flow of a fluid of material particles in a confined space under the influence of a pressure difference\cite{landauhydro}, Eq. (\ref{eqn:poisoille1}) describes the Poiseuille flow of a phonon gas\cite{sussmann, gurzhi,guyer1}. To be specific we consider a 2D hexagonal crystal with rectangular shape of length $L \to \infty$ and width $w$. A constant temperature gradient is applied along $L$ taken as $\vec{x}$ direction. Then $\vec{a}$ is solely a function of $y$ and $\pi^\prime_{ki,ki}$ reduces to $\pi^\prime_{12,12} = \pi^\prime_{66}$. We use Voigt's notation with $(11) \equiv 1$, $(22) \equiv 2$ and $(12) \equiv 6$. We then solve the differential equation (\ref{eqn:poisoille1}) with boundary conditions $\vec{a} (y = \pm w/2) = 0$. Finally we average over the width $w$ and obtain
\begin{align}
\vec{a} =  - \frac{ \langle \omega \vert k_1 v_1 \rangle }{ T \langle k_1 \vert k_1 \rangle} \frac{ 1 }{\omega_V} \big[ 1 - \frac{2}{w \alpha_w} \textrm{tanh} (\frac{\alpha_w w}{2}) \big]        \vec{\nabla} \theta \; ,
\label{eqn:ajpoiseuille}
\end{align}
where $\alpha_w = \sqrt{\omega_V / \pi^\prime_{66}}$ has the dimension of inverse length. Using dimensional arguments we identify $c/ \omega_V$ and $\pi^\prime_{66}/c$, where $c$ is of the order of the sound velocity, with the mean free paths $l_V$ and $l_N$ for momentum dissipating processes and N-processes respectively. These definitions were previously introduced by Gurzhi\cite{gurzhi}. Noting that the corresponding heat current per unit cell reads
\begin{align}
\vec{Q} =  \frac{1}{N} \sum_{\vec{k}, \alpha} \hbar^2 \omega(\vec{k}, \alpha) \vec{v}(\vec{k}, \alpha) n^\prime (\vec{k}, \alpha) \vec{k} \cdot \vec{a} \; ,
\label{eqn:heatcurrectpoiseuille}
\end{align}
we define the thermal conductivity $\kappa_{2D}$ per unit cell from comparison of Eqs. (\ref{eqn:ajpoiseuille}) and (\ref{eqn:heatcurrectpoiseuille}) as 
\begin{align}
\vec{Q} =  - \kappa_{2D} \vec{\nabla} \theta \; .
\label{eqn:fourierbasiclaw}
\end{align}

The thermal conductivity per unit volume\cite{balandin1,seol} $\kappa$ is obtained by division of $\kappa_{2D}$ by the effective volume $v_{3D} = v_{2D} \times h$, where $h$ is the interlayer distance in the corresponding 3D material:
\begin{align}
\kappa =  \frac{ \hbar^2 \langle \omega \vert k_1 v_1 \rangle^2 }{v_{3D} T \langle k_1 \vert k_1 \rangle} \frac{ 1 }{\omega_V} \big[ 1 - \frac{2}{w \alpha_w} \textrm{tanh} (\frac{\alpha_w w}{2}) \big] \; .
\label{eqn:kappapoiseuille}
\end{align}
This expression provides a unified description of the temperature variation of the thermal conductivity. We notice that $\alpha_w$ is determined by both momentum dissipating processes and by N-processes through $\omega_V$ and $\pi^\prime_{66}$ respectively. Anticipating the results of Secs. (\ref{sec:transpcoe}) and (\ref{sec:numresults}) where we show that $\pi^\prime_{66}$ is a strongly increasing function with decreasing temperature, and assuming a sample of finite width such that $\alpha_w w < 2$, we see that $\alpha_w \to 0$ at low $T$. Then expansion of $\textrm{tanh} (\frac{\alpha_w w}{2})$ in Eq. (\ref{eqn:kappapoiseuille}) leads to
\begin{align}
\kappa =  \frac{ \hbar^2 \langle \omega \vert k_1 v_1 \rangle^2 w^2 }{12 v_{3D} T \langle k_1 \vert k_1 \rangle \pi^\prime_{66}}  \; .
\label{eqn:kappapoiseuille2}
\end{align}
Here $\kappa$ is independent of U-processes and intrinsic impurity scattering but limited by the scattering of N-processes against the sample boundaries. At high $T$ where $\alpha_w w >> 2$, Eq. (\ref{eqn:kappapoiseuille}) gives
\begin{align}
\kappa =  \frac{ \hbar^2 \langle \omega \vert k_1 v_1 \rangle^2 }{v_{3D} T \langle k_1 \vert k_1 \rangle \omega_V}  \; .
\label{eqn:kappapoiseuille3}
\end{align}
A quantitative discussion of $\kappa$ as a function of $T$ requires the explicit knowledge of $\omega_V$ and $\pi^\prime_{66}$, and will be given in Sec. \ref{sec:numresults}. 

We observe that the results of this subsection have been obtained for $\omega_V < \omega^l$. At high $T$ or with large impurity scattering this condition breaks down. At low frequencies and long wavelengths such that $\omega$ and $q^2 \pi^\prime$ can be neglected in comparison with $\omega_V$ in the denominator of Eq. (\ref{eqn:ajdefdisplaceddif}), we find that Eq. (\ref{eqn:nu0flexcomplete2}) reduces to Eq. (\ref{eqn:nuflexcomplete}) with $\lambda$ given by Eq.(\ref{eqn:lambdatensordef}), i.e. we recover the situation of subsection \ref{subsec:energyconservation}.

We conclude by showing how to calculate $\omega_V$ without impurity scattering. Starting from Eq. (\ref{eqn:collisionextectedvalue}) with $\nu (\vec{k},\alpha) = k_i$ we obtain for the relaxation frequency $\omega_U$ due to U-processes:
\begin{align}
\omega_U &= \frac{\langle k_i \vert \hat{C} \vert k_i \rangle}{\langle k_i \vert k_i \rangle} \notag\\
&=\frac{1}{N \langle k_i \vert k_i \rangle} \sum_{\vec{k}\vec{p}\vec{h},j} \bigg\{ W \left( \! \begin{array}{ccc} \zeta & j & \zeta \\ \vec{k} & \vec{p} & \vec{h} \end{array} \! \right) \big[ k_i - h_i - p_i \big]^2  \notag\\
    &+\frac{1}{2} W \left( \! \begin{array}{ccc} j & \zeta & \zeta \\ \vec{p} & \vec{h} & \vec{k} \end{array} \! \right)  \big[ k_i + h_i - p_i \big]^2 \bigg\} \; \notag.
\end{align}
Symmetrizing the first term on the right hand side with respect to $(\vec{p},j)$ and $(\vec{h},\zeta)$ and re ordering terms,
\begin{align}
\omega_U &= \frac{1}{2N \langle k_i \vert k_i \rangle} \sum_{\vec{k}\vec{p}\vec{h},j} \big[ k_i + h_i + p_i \big]^2
  \notag\\
 &\times \Big\{ 2 W \left( \! \begin{array}{ccc} \zeta & j & \zeta \\ \vec{k} & \text{-}\vec{p} & \text{-}\vec{h} \end{array} \! \right) + W \left( \! \begin{array}{ccc} j & \zeta & \zeta \\ \text{-}\vec{p} & \vec{h} & \vec{k} \end{array} \! \right)  \Big\} \; .
\label{eqn:omega_U_calc_2}
\end{align}
Typical calculations of thermal conductivity by \textit{ab initio} methods\cite{lee_naturecom,cepe_naturecom,xunhydro} involve also  all combinations of (only) in-plane scattering processes. For a quantitative comparison, the inclusion of in-plane scattering to $\omega_U$ can be done by adding to the right hand side of Eq. (\ref{eqn:omega_U_calc_2}) the terms
\begin{align}
    W \left( \! \begin{array}{ccc} i & j & k \\ \vec{k} & \vec{p} & \vec{h} \end{array} \! \right) &= 2 \pi \hbar \; \bigg\vert \Psi^{(3)} \left(\! \begin{array}{ccc} i & j & k \\ -\vec{k} & \vec{p} & \vec{h} \end{array} \! \right) \bigg\vert^2 \notag\\
    &\times \sqrt{k_B T n^{\prime}(\vec{k},i)n^{\prime}(\vec{p},j)n^{\prime}(\vec{h},k)} \notag\\
    &\times \delta(\omega(\vec{p},j)-\omega(\vec{k},i)+\omega(\vec{h},k)) \; , 
\label{eqn:defomegainplane}
\end{align}
where
\begin{align}
    \Psi^{(3)} \left(\! \begin{array}{ccc} i & j & k \\ \vec{k} & \vec{p} & \vec{h} \end{array} \! \right)= \frac{\Phi^{(3)} \left(\! \begin{array}{ccc} i & j & k \\ \vec{k} & \vec{p} & \vec{h} \end{array} \! \right)}{\sqrt{8 \omega(\vec{k},i) \omega(\vec{p},j) \omega(\vec{h},k)}} \; . 
\label{eqn:defpsimayusinplane}
\end{align}
With this addition the relaxation frequency $\omega_U$ will be referred as $\tilde{\omega}_U$ instead. The expression for $\Phi^{(3)} (\vec{k},i;\vec{p},j;\vec{h},k)$ is similar to Eq. (\ref{eqn:anharmonichamiltonian}), except for the interatomic potentials $\phi^{(3)}_{ijk} (A_1;B_s)$, that refer to in-plane anharmonic interactions. For more details on the specific values and symmetries see Ref. (\onlinecite{seba}).

%
\section{Temperature and Displacement Response}
\label{sec:respfunc}
%

Having obtained the hydrodynamic solutions to Eqs. (\ref{eqn:soundwaves}) and (\ref{eqn:nuflex}), we now investigate their implications on the structure of the correlations functions. Such a study is also of relevance in view of experiments\cite{griffin}.

In the following we will restrict ourselves again to 2D crystals of hexagonal symmetry. Then there are only two independent elastic constants $C_{11,11}$ and $C_{11,22}$, while $2 C_{12,12} = C_{11,11} - C_{11,22}$. Using Voigt's notation and observing that in 2D crystals these quantities have the dimension of tension coefficients, we write $\gamma_{11} \equiv C_{11,11}$, $\gamma_{12} \equiv C_{11,22}$, and $\gamma_{66} \equiv C_{12,12}$. The other fourth rank tensors $\eta_{ij,kl}$ and $\pi^{\prime}_{ij,kl}$ also have the same symmetries as the elastic constants. In order to calculate the dynamic structure functions $S_{\gamma} (\vec{q}, \omega)$ with $\gamma = \{ss, \theta \theta \}$, we first determine the dynamic susceptibilities $\chi_{\gamma} (\vec{q}, \omega)$, take the imaginary parts $\chi^{\prime \prime}_{\gamma} (\vec{q}, \omega)$ and use the fluctuation dissipation theorem\cite{callen,kadanoff}
\begin{align}
    S_\gamma (\vec{q}, \omega) = [1+ n(\omega)] \chi^{\prime \prime}_\gamma (\vec{q}, \omega) \; ,
\label{eqn:structurefactor}
\end{align}
with $n (\omega) = [\textrm{exp}(\hbar \omega / k_B T) -1]^{-1}$. In accordance with subsections \ref{subsec:energyconservation} and \ref{subsec:energyandmomentumconservation} we will consider separately the diffusive regime and the second sound regime.

%
\subsection{Diffusive regime}
\label{subsec:difreg}
%

In order to calculate the displacement-displacement response function $S_{ss}$ we consider Eqs. (\ref{eqn:soundwaves3}) and (\ref{eqn:nuflexcomplete}) in the absence of the external temperature source $\Theta (\vec{q}, \omega)$. Taking $\vec{q} = \{ q_1, 0 \}$ we solve Eq. (\ref{eqn:nuflexcomplete}) with respect to $\theta (\vec{q}, \omega)$ and then substitute the result into Eq. (\ref{eqn:soundwaves3}). For the longitudinal case $\vec{s} = \{ s_1, 0\}$ we obtain:
\begin{align}
    &\Big[ \omega^2  - c_L^2 q_1^2 +  i q_1^2 \omega \hat{\eta}_{11}  - q_1^2 \frac{\omega \beta^{\prime 2}}{\omega + i q_1^2 \lambda} \Big] s_1(\vec{q},\omega) \notag \\
    &= - F_1 (\vec{q},\omega) /m \; .
\label{eqn:suscp1}
\end{align}
Here we have defined 
\begin{align}
    c_L^2 &= \gamma_{11} v_{2D} / m \; , \notag \\
    \beta^{\prime 2} &= \frac{v_{2D}^2 T \beta^2}{m c_v} \; , \notag\\
    \hat{\eta}_{11} &= \frac{v_{2D}}{m} \eta_{11,11} \; ,
\label{eqn:defcl}
\end{align}
where $c_L$ is the isothermal sound velocity and $\beta^{\prime 2}$ accounts for the coupling to local temperature fluctuations. Eq. (\ref{eqn:suscp1}) describes damped longitudinal sound waves. Beside the kinematic viscosity $\hat{\eta}_{11}$, temperature diffusion acts as a mechanism of sound absorption.

\begin{figure}
\includegraphics[trim = 0.5cm 0cm 0cm 0.0cm, clip, width=8.7cm]{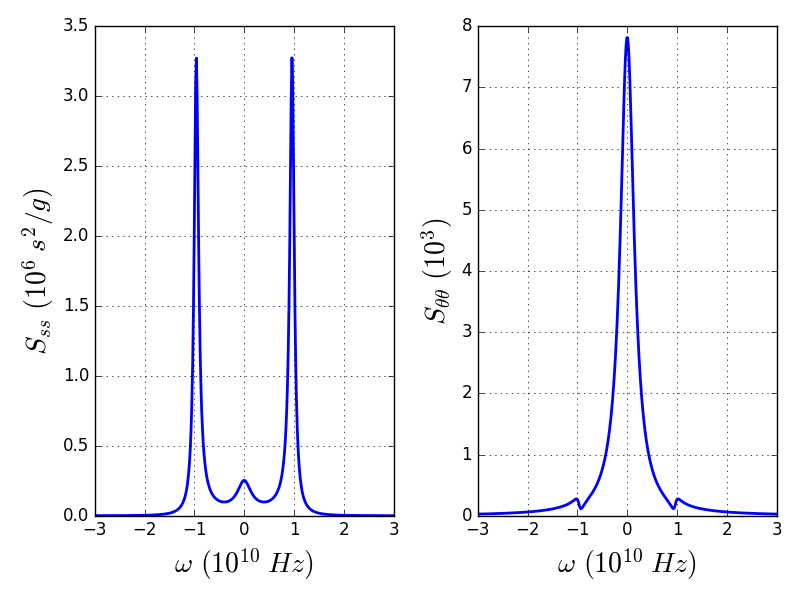}
\caption{Left panel: Plot of the dynamic structure function $S_{ss} (\vec{q}, \omega)$ in the diffusive regime according to Eq. (\ref{eqn:suslongitudinal}). Right panel: Plot of the dynamic structure function $S_{\theta \theta} (\vec{q}, \omega)$ (dimentionless) in the diffusive regime according to Eq. (\ref{eqn:sustemp1}). All relevant parameters are given in Table. \ref{tab1}.}
\label{fig:res_disp}
\end{figure}

Differentiating Eq. (\ref{eqn:suscp1}) on both sides with respect to the external force $F_1 (\vec{q},\omega)$ we obtain the longitudinal displacement-displacement susceptibility
\begin{align}
    \chi_{ss}(\vec{q},\omega) &= \frac{\delta s_1(\vec{q},\omega)}{\delta F_1(\vec{q},\omega)} =  \frac{-1}{m}   \notag\\
    &\bigg\{ \omega^2 - c_L^2 q_1^2 +  i q_1^2 \omega \hat{\eta}_{11} - q_1^2 \frac{\omega \beta^{\prime 2}}{\omega + i q_1^2 \lambda} \bigg\}^{-1} \; .
\label{eqn:suslongitudinal}
\end{align}
The resonances can be analyzed by hand in important limiting cases. At large sound wave frequencies where temperature fluctuations can not follow, i. e. $\omega = c_L q_1 > q_1^2 \lambda$, the heat diffusion pole at $\omega = -i q_1^2 \lambda$ can be neglected and the resonances are damped sound waves centered at $\pm c_L^{ad} q_1$, where
\begin{align}
    &c_L^{ad} = \big( c_L^2 + \beta^{\prime 2} \big)^{\frac{1}{2}} \; ,
\label{eqn:velsoundadiab}
\end{align}
is the adiabatic sound velocity. In the opposite regime, i. e. $\omega < q_1^2 \lambda$, the heat diffusion pole leads to an additional resonance at $\omega = 0$, the Landau-Placzek peak\cite{bls}. The relative strength of these resonances is most conveniently studied by observing that the denominator of Eq. (\ref{eqn:suslongitudinal}) is a polynomial of third order in $\omega$. Finding the roots by performing a perturbation expansion in powers of small $q_1$ and making subsequently a partial fraction decomposition we obtain 
\begin{align}
    &\chi_{ss}(\vec{q},\omega) = \frac{1}{m}  \bigg\{ \frac{1}{2 c_L^{ad} q_1} \big[   \frac{1}{\omega + c_L^{ad} q_1 + i q_1^2 \hat{\eta}_{11} / 2} \notag\\
    &- \frac{1}{\omega - c_L^{ad} q_1 + i q_1^2 \hat{\eta}_{11} / 2} \big] + \frac{i \lambda \beta^{\prime 2}}{ {c_L^{ad}}^4 ( \omega + i q_1^2 \lambda (\frac{c_L}{c_L^{ad}})^2 )} \bigg\} \; .
\label{eqn:suslongitudinalpartialdec}
\end{align}
The first two terms on the right hand side of Eq. (\ref{eqn:suslongitudinalpartialdec}) reflect the sound wave resonances while the third term is due to temperature diffusion. We next determine the imaginary part $\chi^{\prime \prime}_{ss}(\vec{q},\omega)$ and calculate the sum rule\cite{GM2}
\begin{align}
    m c_L^2 q_1^2 \int_{-\infty}^{\infty} \frac{\textrm{d} \omega}{\pi} \frac{\chi^{\prime \prime}_{ss}(\vec{q},\omega)}{\omega} = (\frac{c_L}{c_L^{ad}})^2 + (\frac{\beta^{\prime}}{c_L^{ad}})^2 \; .
\label{eqn:sumrulesusceptibility}
\end{align}
The first term on the right hand side has its origin in the Brillouin doublet and the second term is due to the heat diffusion peak. We see that as a consequence of the coupling of lattice deformations to temperature diffusion, the elastic sum rule is not completely  exhausted by the isothermal sound waves but there is in addition a contribution of weight $(\beta^{\prime} / c_L^{ad})^2$ due to the central Landau-Placzek peak. In the left panel of Fig. \ref{fig:res_disp} we have plotted the dynamic structure function $S_{ss} (\vec{q}, \omega)$ for a hypothetical 2D crystal with the parameters listed in Table \ref{tab1}. 
\begin{table}
\begin{center}
\begin{tabular}{ |c|c|c|c|c|c|c| }
 \hline
 $m$ & $q_1$ (m$^{-1}$) & $c_L$ (ms$^{-1}$) & $\beta^{\prime 2}$ & $q_1 \hat{\eta}_{11}$ & $q_1 \lambda$  & $T$ (K)\\ 
 \hline
 $2 \: m_C$ & $4 \times 10^5$  & $23 \times 10^3$ & $c_L^2 / 10$ & $c_L / 10$ & $c_L / 5$ & $100$ \\
 \hline
\end{tabular}
\end{center}
\caption{Parameters used for the calculation of the structure functions $S_{ss}$ and $S_{\theta \theta}$ in the diffusive regime.}
\label{tab1}
\end{table}
As we can observe, the Brillouin doublet and the Landau-Placzek peak are clearly present. 

In order to calculate the dynamic temperature-temperature susceptibility, we put $F_i (\vec{q}, \omega) = 0$ in Eq. (\ref{eqn:soundwaves3}) and eliminate $s_j$ from the coupled system of Eqs. (\ref{eqn:soundwaves3}) and (\ref{eqn:nuflexcomplete}). Differentiating the resulting equation for $\theta (\vec{q}, \omega)$ with respect to the external temperature source $\Theta (\vec{q}, \omega)$ we obtain the susceptibility
\begin{align}
    &\chi_{\theta \theta}(\vec{q},\omega) = \frac{\delta \theta(\vec{q},\omega)}{\delta \Theta(\vec{q},\omega)} = \frac{i q_1^2 \lambda}{\omega + i q_1^2 \lambda - Q (q_1, \omega, T)} \; ,
\label{eqn:sustemp1}
\end{align}
with
\begin{align}
    Q (\vec{q}, \omega) = \frac{q_1^2 \omega \beta^{\prime 2}}{ \omega^2 -  c_L^2 q_1^2 + i \omega q_1^2 \hat{\eta}_{11} } \; . \notag
\end{align}
Repeating the same steps made above for $\chi_{ss}(\vec{q}, \omega)$, we obtain again three resonances: a sound wave doublet at $\pm c_L^{ad} q_1$ and a heat diffusion peak at $\omega = 0$. The difference with $\chi_{ss}(\vec{q}, \omega)$ is that the strength (weight) of the resonances has been reversed: here the heat diffusion peak is the dominating feature, while the sound wave doublet has weight $(\beta^{\prime} / c_L^{ad})^2$. The corresponding scattering law $S_{\theta \theta} (\vec{q}, \omega)$ is again readily obtained by taking the imaginary part of Eq. (\ref{eqn:sustemp1}) and applying relation (\ref{eqn:structurefactor}). The result is plotted in the right panel of Fig. \ref{fig:res_disp} where again we have used the parameters listed in Table \ref{tab1}.

%
\subsection{Second sound regime}
\label{subsec:ssreg}
%

Here we start from the wave equation (\ref{eqn:nu0flexcomplete5approx2secondsound}) for local temperature fluctuations. As already noticed at the end of Sec. \ref{sec:solutions}, the wave equation for the lattice deformations is formally the same as Eq. (\ref{eqn:soundwaves3}). We follow the same steps as subsection \ref{subsec:difreg}, now applied to Eqs. (\ref{eqn:soundwaves3}) and (\ref{eqn:nu0flexcomplete5approx2secondsound}). The longitudinal displacement-displacement susceptibility is readily obtained as
\begin{align}
    \chi_{ss}(\vec{q},\omega) &= \frac{\delta s_1(\vec{q},\omega)}{\delta F_1(\vec{q},\omega)} =   -\frac{1}{m} \bigg\{  \omega^2 - c_L^2 q_1^2 \notag\\
    &+ i q_1^2 \omega \hat{\eta}_{11} - \omega^2 q_1^2 \beta^{\prime 2} R(\vec{q}, \omega) \bigg\}^{-1} \; ,
\label{eqn:sussecondlongitudinal}
\end{align}
where
\begin{align}
    R(\vec{q}, \omega) &= \frac{1}{\omega^2 - q_1^2 V^2_\theta + i \omega q_1^2 \tilde{\lambda} } \; , \notag\\
    \tilde{\lambda} &= \frac{\omega_V}{q_1^2} +  \lambda^\prime + \pi^\prime_{11} \; .
\end{align}
The resonances of $\chi_{ss}(\vec{q},\omega)$ are now given by the zeros of a polynomial of fourth order in $\omega$. In the limit of small wave vectors we apply again perturbation theory and perform a partial fraction decomposition. The result reads
\begin{align}
    \chi_{ss}(\vec{q},\omega) &= -\frac{1}{m} \bigg\{  
\frac{1}{2 \tilde{c}_L q_1} \big( 1-\frac{\beta^{\prime 2} V^2_\theta}{(c_L^2 - V^2_\theta)^2} \big) \times \notag\\
&\big[ \frac{1}{\omega - \tilde{c}_L q_1 + i q_1^2 \hat{\eta}_{11} /2 } - \frac{1}{\omega + \tilde{c}_L q_1 + i q_1^2 \hat{\eta}_{11} /2} \big]  \notag\\
&+ \frac{\beta^{\prime 2} V^2_\theta}{2 \tilde{V} q_1 (c_L^2 - V_\theta^2)^2} \times \notag\\
&\big[ \frac{1}{\omega - \tilde{V}_\theta q_1 + i q_1^2 \tilde{\lambda} /2} - \frac{1}{\omega + \tilde{V}_\theta q_1 + i q_1^2 \tilde{\lambda} /2} \big] \bigg\} \; ,    
\label{eqn:sussecondlongitudinalpartial}
\end{align}
with
\begin{align}
    \tilde{c}_L &= \Big[ c_L^2 +  \frac{\beta^{\prime 2} c_L^2}{c_L^2 - V^2_\theta} \Big]^{\frac{1}{2}} \; , \notag\\
    \tilde{V}_\theta &= \Big[ V^2_\theta +  \frac{\beta^{\prime 2} V^2_\theta}{V^2_\theta - c_L^2} \Big]^{\frac{1}{2}} \; .   
\label{eqn:deftildevars}
\end{align}
\begin{table}
\begin{center}
\begin{tabular}{ |c|c|c|c|c|c|c|c| }
 \hline
 $m$ & $q_1$ (m$^{-1}$) & $c_L$ (ms$^{-1}$) & $\beta^{\prime 2}$ & $q_1 \hat{\eta}_{11}$ & $q_1 \tilde{\lambda}$  & $V$ & $T$ (K)\\ 
 \hline
 $2 \: m_C$ & $4 \times 10^5$  & $23 \times 10^3$ & $c_L^2 / 10$ & $c_L / 10$ & $c_L / 7$ & $c_L / 2$ & $100$ \\
 \hline
\end{tabular}
\end{center}
\caption{Parameters used for the calculation of the structure functions $S_{ss}$ and $S_{\theta \theta}$ in the second sound regime.}
\label{tab2}
\end{table}
\begin{figure}
\includegraphics[trim = 0.5cm 0cm 0cm 0.0cm, clip, width=8.7cm]{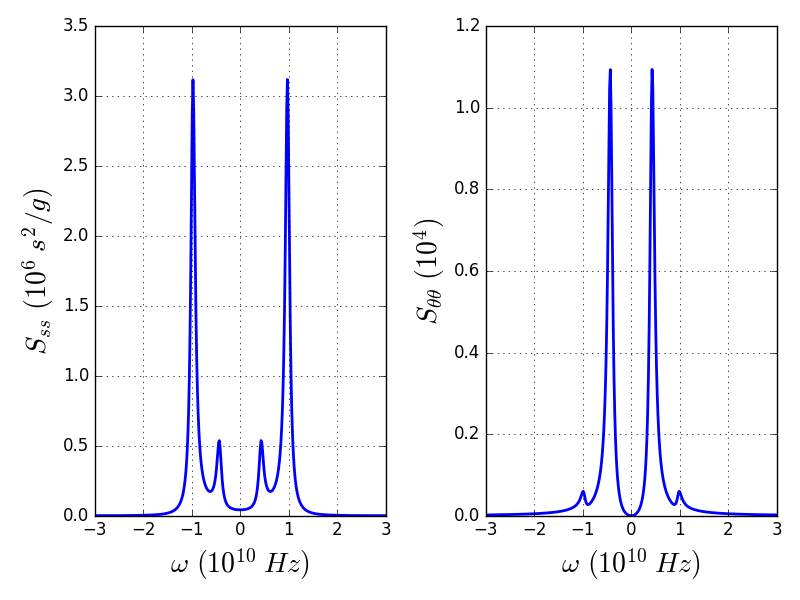}
\caption{Left panel: Plot of the dynamic structure function $S_{ss} (\vec{q}, \omega)$  in the second sound regime according to Eq. (\ref{eqn:sussecondlongitudinal}). Right panel: Plot of the dynamic structure function $S_{\theta \theta} (\vec{q}, \omega)$  in the second sound regime according to Eq. (\ref{eqn:sussecondtemp}). All relevant parameters are given in Table. (\ref{tab2}).}
\label{fig:res_temp}
\end{figure}
The symbol $\tilde{c}_L$ stands for the first sound velocity renormalized by thermal tension due to heat waves (second sound), while $\tilde{V}_\theta$ is the second sound velocity renormalized by the presence of coherent dynamic lattice deformations (first sound waves). From Eq. (\ref{eqn:sussecondlongitudinalpartial}) we see that the resonances are given by a sound wave doublet centered at $\pm \tilde{c}_L q_1$ and a second sound doublet at $\pm \tilde{V}_\theta q_1$. The weight of these resonances in the dynamic displacement-displacement susceptibility or equivalently in the dynamic structure function $S_{ss} (\vec{q}, \omega)$ is given by $1 - \beta^{\prime 2} V^2_\theta / (c_L^2 - V^2_\theta)^2$ and $\beta^{\prime 2} V^2_\theta / (c_L^2 - V^2_\theta)^2$, respectively. In the right panel of Fig. \ref{fig:res_temp} we have plotted the dynamic structure function $S_{ss} (\vec{q}, \omega)$ for a 2D crystal in the second sound regime with the parameters listed in Table (\ref{tab2}). According to the partial fraction decomposition in Eq. (\ref{eqn:sussecondlongitudinalpartial}) there are four distinguishable peaks. The doublet with larger intensity appears at $\pm \tilde{c}_L q_1 \approx \pm 1.06 \: c_L q_1$ while the smaller one, closer to $\omega = 0$, appears at $\pm \tilde{V}_\theta q_1 \approx \pm 0.93 \: V_\theta q_1$. As we can see, the renormalization barely changes the position of the first and second sound peaks (with the a priori chosen parameters) and so the main difference with the diffusive regime is the splitting of the central Landau-Placzek peak into a second sound doublet.

The dynamic temperature-temperature susceptibility is obtained by differentiating  Eq. (\ref{eqn:nu0flexcomplete5approx2secondsound}) for $\theta (\vec{q}, \omega)$ with respect to the external temperature source $\Theta (\vec{q}, \omega)$, after having eliminated $s (\vec{q}, \omega)$ by means of Eq. (\ref{eqn:soundwaves3}). The result reads 
\begin{align}
    &\chi_{\theta \theta}(\vec{q},\omega) = \frac{\delta \theta(\vec{q},\omega)}{\delta \Theta(\vec{q},\omega)} \notag\\
    &=  \frac{-q_1^2 V_{\theta}^2 + i \omega q_1^2 \lambda^\prime}{\omega^2 - q_1^2 V^2_\theta +i \omega q_1^2 \tilde{\lambda}  - \omega^2 q_1^2 \beta^{\prime 2} P(\vec{q}, \omega)} \; ,
\label{eqn:sussecondtemp}
\end{align}
where
\begin{align}
    &P(\vec{q}, \omega) = \frac{1}{\omega^2 - q_1^2 c_L^2 + i \omega q_1^2 \hat{\eta}_{11}} \; . \notag
\end{align}
The resonances of $\chi_{\theta \theta}(\vec{q},\omega)$ are given by the zeros of the same polynomial of fourth order in $\omega$ as in the case of $\chi_{ss} (\vec{q}, \omega)$. Carrying out again a partial fraction decomposition we obtain
\begin{align}
    \chi_{\theta \theta}(\vec{q},\omega) &= -q_1^2 V^2_\theta \bigg\{  
\frac{\beta^{\prime 2} c_L^2}{2 \tilde{c}_L q_1 (c_L^2 - V^2_\theta)^2} \notag\\
&\big[ \frac{1}{\omega - \tilde{c}_L q_1 + i q_1^2 \hat{\eta}_{11} /2 } - \frac{1}{\omega + \tilde{c}_L q_1 + i q_1^2 \hat{\eta}_{11} /2} \big]  \notag\\
&+ \frac{1}{2 \tilde{V}_\theta q_1} \big(1- \frac{\beta^{\prime 2} c_L^2}{(c_L^2 - V^2_\theta)^2} \big) \times \notag\\
&\big[ \frac{1}{\omega - \tilde{V}_\theta q_1 + i q_1^2 \tilde{\lambda} / 2} - \frac{1}{\omega + \tilde{V}_\theta q_1 + i q_1^2 \tilde{\lambda} / 2} \big] \bigg\} \; .    
\label{eqn:sussecondtemppartial}
\end{align}
Comparison with Eq. (\ref{eqn:sussecondlongitudinalpartial}) for $\chi_{ss}$ shows that the (first) sound doublet has now the smaller weight $\beta^{\prime 2} c_L^2 / (c_L^2 - V^2)^2$ while the second sound doublet has the larger weight $1 - \beta^{\prime 2} c_L^2 / (c_L^2 - V^2)^2$. This trend can be observed clearly in the right panel of Fig. \ref{fig:res_temp} where we plot the values of $\chi_{\theta \theta}(\vec{q},\omega)$ in the second sound regime for the same parameters listed in Table. \ref{tab2}.

Here a remark on the $T$-dependence of the thermal tension $\beta$ is on order. At low $T$ where flexural modes are dominant, the thermal expansion $\alpha_T$ or equivalently $\beta$ is negative in 2D crystals\cite{liftshitz, mounet} as has been found by experiments in graphene\cite{yoon, bao}. At higher $T$ the excitation of in-plane phonons gives in addition a positive contribution to $\alpha_T$ and $\beta$ which even can lead to a change of sign as is suggested by atomistic Monte Carlo simulations\cite{zakhar}. This effect is material\cite{sevic} and size dependent\cite{seba}.

%
\section{Transport coefficients}
\label{sec:transpcoe}
%

In Sec. \ref{sec:solutions} we derived a series of hydrodynamic equations by applying a formal solution method to the kinetic equations. Thereby the non-equilibrium phonon density has been expanded in terms of eigenfunctions of the collision operator $\hat{C}$. The corresponding eigenvalues are the relaxation frequencies of the crystal. Separating the spectrum into two classes corresponding to slowly varying secular variables (conserved and quasi conserved quantities) and fast varying variables respectively, we have used perturbation theory to formulate the hydrodynamic equations for the secular variables such as lattice deformations and local temperature. In addition to restoring forces that account for the oscillatory part (e.g. elastic constants and sound velocities), the hydrodynamic equations contain dissipative terms that account for the transformation of kinetic energy into heat. The dissipative terms are characterized by the so-called kinetic coefficients or transport coefficients such as viscosity and thermal conductivity. 

The transport coefficients $\eta$, $\lambda$, $\lambda^\prime$ and $\pi^\prime$ have been formulated in terms of projections of the currents of secular variables onto the eigenfunctions $\chi^l$ belonging to the fast variables of the collision operator. While formally compact, expressions (\ref{eqn:viscositytensordef}), (\ref{eqn:lambdatensordef}) and (\ref{eqn:piprimedef}) are not very practical, since neither $\chi^l$ nor the corresponding eigenfrequencies $\omega^l$ are explicitly known. Here we will follow a more systematic approach where transport coefficients are written as correlation functions of currents of secular variables\cite{Kubo,mori} in the limit of zero frequency\cite{kadanoff}. Taking in turn the currents as secular variables, G{\"o}tze and one of the present authors\cite{GM3} have expressed the correlation functions in terms of memory kernels for which closed expressions in form of multiple integrals over the Brillouin zone are obtained.

As shown in Ref. \onlinecite{GM3}, a transport coefficient $\Lambda$ that describes the decay of a secular variable $A$ (or a set of secular variables) can be cast into the form of a matrix equation
\begin{align}
\Lambda = i \chi^{JJ} (0) (\hat{m}^{KK} (i0))^{-1} \chi^{JJ} (0) (\chi^{AA}(0))^{-1} \; ,
\label{eqn:biglambdadef}
\end{align}
where the factors on the right hand side can be calculated by perturbation theory. The tensor properties of the factors depend on the nature of the transport coefficient. Here $\chi^{JJ} (0)$ is the static current-current susceptibility, $J$ stands for the non-secular part of the current of $A$, and  $\chi^{AA} (0)$ is the static $A$-$A$ susceptibility. The memory function $\hat{m}^{KK} (i0)$ is the zero frequency limit $\omega \to +i0$ of a higher hierarchy current-current correlation function, where the current $K$ is obtained from $J$ taken in turn as a secular variable. Closed expressions of the memory function are obtained by evaluating the K-K correlation functions by means of decoupling techniques\cite{zubarev,GM3} and by taking subsequently the zero frequency limit\cite{kadanoff}. See Eqs. (\ref{eqn:matrixm1}), (\ref{eqn:matrixm2}) and (\ref{eqn:matrixm3}) below. 

In the following, we will present the main steps of implementation of Eq. (\ref{eqn:biglambdadef}) for the calculation of the transport coefficients $\hat{\eta}_{11}$, $\lambda$, $\lambda^\prime$ and $\pi^{\prime}_{66}$. For additional details of the method, see Ref. (\onlinecite{GM3}). In order to calculate the in-plane kinematic viscosity $\hat{\eta}_{11}$, we take the in-plane momentum $\vec{p} (\vec{q})$, Eq. (\ref{eqn:centerofmass2}) conjugate to the in-plane deformation $\vec{s} (\vec{q})$ as secular variable. Then Eq. (\ref{eqn:biglambdadef}) leads to
\begin{align}
\hat{\eta}_{pq} = i \chi^{JJ}_{pn} (0) (\hat{m}^{KK} (i0))^{-1}_{nl} \chi^{JJ}_{lq} (0) (\chi^{pp} (0))^{-1} \; .
\label{eqn:etatransportdef}
\end{align}
The indexes $p$, $q$, $n$ and $l$ here correspond to Voigt's notation: $\hat{\eta}_{11,11} \equiv \hat{\eta}_{11}$, $\hat{\eta}_{22,22} \equiv \hat{\eta}_{22}$, $\hat{\eta}_{11,22} \equiv \hat{\eta}_{12}$ and $\hat{\eta}_{12,12} \equiv \hat{\eta}_{66}$. These definitions apply to all fourth rank tensors, including current-current susceptibilities and memory kernels as well. We now specify all factors that occur in $\eta_{pq}$. The momentum-momentum static susceptibility is given by $\chi_1^{pp} (0) = 1$. The non secular part of the current of $\vec{p}(\vec{q})$ reads
\begin{align}
J_n = \frac{1}{\sqrt{N}} \sum_{\vec{k}} \gamma_n (\vec{k}, \zeta) b^{\zeta \dagger}_{\vec{k}} b^{\zeta}_{\vec{k}} \; ,
\label{eqn:cosarara}
\end{align}
where
\begin{align}
\gamma_n (\vec{k}, \zeta)  = \frac{\hbar}{\sqrt{m}} [h_n (\vec{k}, \zeta) - \omega (\vec{k}, \zeta) \frac{ \beta v_{2D} }{c_v} ]  \; ,
\label{eqn:gammadef1}
\end{align}
with\cite{SMP}
\begin{align}
h_1 (\vec{k}, \zeta)  \equiv h_{11} (\vec{k}, \zeta)  &= \frac{\sqrt{3} h^{(3)} a^3}{16 m \omega (\vec{k}, \zeta)} (k_x^2 + \frac{k_y^2}{3}) \; , \notag\\
h_2 (\vec{k}, \zeta)  \equiv h_{22} (\vec{k}, \zeta)  &= \frac{\sqrt{3} h^{(3)} a^3}{16 m \omega (\vec{k}, \zeta)} (k_y^2 + \frac{k_x^2}{3}) \; , \notag\\
h_6 (\vec{k}, \zeta)  \equiv h_{12} (\vec{k}, \zeta)  &= \frac{ h^{(3)} a^3}{8 \sqrt{3}  m \omega (\vec{k}, \zeta)} k_x k_y  \; .
\label{eqn:hdeftransport}
\end{align}
Here $h^{(3)}$ is the anharmonic coupling constant\cite{seba} between in-plane and out-of-plane displacements, and $a$ is the lattice constant. The current-current susceptibility reads
\begin{align}
\chi_{pn}^{JJ} (0) = \frac{1}{N} \sum_{\vec{k}} \gamma_p (\vec{k}, \zeta) n^\prime (\vec{k}, \zeta) \gamma_n (\vec{k}, \zeta) \; .
\label{eqn:chijjdef1}
\end{align}
Symmetry implies $\chi_{11}^{JJ} (0) = \chi_{22}^{JJ} (0)$ and $\chi_{12}^{JJ} (0) = \chi_{21}^{JJ} (0)$. The subtraction of the term proportional to $\hbar \omega(\vec{k},\zeta)$ on the right hand side of Eq. (\ref{eqn:gammadef1}) corresponds to the absence of the $l = 0$ term in Eq. (\ref{eqn:viscositytensordef}). The elements of the matrix $\hat{m}^{KK} (i0)$ are
\begin{align}
&\hat{m}^{KK}_{nl} (i0)  = \frac{i \pi \hbar}{N} \sum_{\vec{k},\vec{k}_1,\vec{k}_2,i} \vert \Psi \left( \! \begin{array}{ccc} i & \zeta & \zeta \\ \vec{k} & \vec{k}_1 & \vec{k}_2 \end{array} \! \right) \vert^2  \times \notag\\
&\sqrt{k_B T n^\prime (\vec{k},i) n^\prime (\vec{k}_1,\zeta) n^\prime (\vec{k}_2,\zeta)} \Big\{ \big[ \gamma_n (\vec{k}_1,\zeta) - \gamma_n (\vec{k}_2,\zeta) \big]   \notag\\
&\times \big[ \gamma_l (\vec{k}_1,\zeta) - \gamma_l (\vec{k}_2,\zeta) \big] 
\big[ \delta (\omega(\vec{k},i) -\omega(\vec{k}_1,\zeta) + \omega(\vec{k}_2,\zeta))  \notag\\
&+ \delta (\omega(\vec{k},i) +\omega(\vec{k}_1,\zeta) - \omega(\vec{k}_2,\zeta)) \big] \notag\\
&+ \big[ \gamma_n (\vec{k}_1,\zeta) + \gamma_n (\vec{k}_2,\zeta) \big] \big[ \gamma_l (\vec{k}_1,\zeta) + \gamma_l (\vec{k}_2,\zeta) \big] \notag\\
&\times \delta (\omega(\vec{k},i) - \omega(\vec{k}_1,\zeta) - \omega(\vec{k}_2,\zeta)) \Big\} \; ,
\label{eqn:matrixm1}
\end{align}
where $\omega(\vec{k}, i)$ with $i=\{ 1,2 \}$ are the in-plane phonon frequencies and $\Psi (\vec{k}, i ; \vec{k}_1, \zeta ; \vec{k}_2 , \zeta)$ has been defined by Eq. (\ref{eqn:defpsimayus}). While a quantitative evaluation of $\hat{\eta}$ and the other kinetic coefficients is only possible by numerical calculations (see next section), the asymptotic behavior as a function of temperature has to be obtained by analytical means. Assuming a linear dispersion for the in-plane phonon frequencies and a quadratic dispersion for the flexural mode, we have studied the $T$-dependence of the various susceptibilities and memory kernels. We then find by means of Eqs. (\ref{eqn:chijjdef1}) and (\ref{eqn:matrixm1}) that in the limit $T \to 0$ the susceptibility $\chi^{JJ} (0)$ diverges while $m^{KK}(i0)$ vanishes $\propto T$. Hence we conclude that $\hat{\eta}$ diverges with $T \to 0$. At high $T$ we find that $\chi^{JJ}(0) \propto T$ and $m^{KK} (i0) \propto T^2$, hence $\hat{\eta}$ tends to a constant value.

Next we consider the thermal diffusion coefficient $\lambda_{ij} = \lambda \delta_{ij}$ which has been introduced in Sec. \ref{sec:solutions}, Eq. (\ref{eqn:lambdatensordef}) The corresponding secular variable in this case is the harmonic phonon energy density
\begin{align}
\epsilon (\vec{q})  = \frac{1}{\sqrt{N}} \sum_{\vec{k}, \alpha} \hbar \omega (\vec{k}, \alpha) \: b^{\alpha \dagger}_{\vec{k} - \frac{\vec{q}}{2}} \: b^{\alpha}_{\vec{k} + \frac{\vec{q}}{2}} \; .
\label{eqn:energydensitytransport}
\end{align}
Expression (\ref{eqn:biglambdadef}) leads to the thermal diffusion coefficient
\begin{align}
\lambda = i \chi^{JJ} (0) (\hat{m}^{KK} (i0))^{-1} \chi^{JJ} (0) (\chi^{\epsilon \epsilon} (0))^{-1} \; ,
\label{eqn:lambdatransportdef}
\end{align}
where all factors on the right hand side are scalars. The energy-energy density susceptibility is given by
\begin{align}
\chi^{\epsilon \epsilon} (0) = \hbar^2 \langle \omega \vert \omega \rangle = T c_v \; ,
\label{eqn:energysustransportdef}
\end{align}
and the corresponding current reads
\begin{align}
J_i = \frac{1}{\sqrt{N}} \sum_{\vec{k},\alpha} \gamma_i (\vec{k}, \alpha) b^{\alpha \dagger}_{\vec{k}} b^{\alpha }_{\vec{k}}  \; ,
\label{eqn:corrientelambda}
\end{align}
with
\begin{align}
\gamma_i (\vec{k}, \alpha) = \hbar \omega (\vec{k}, \alpha) v_i (\vec{k}, \alpha) \; .
\label{eqn:lambdacurrenttransport}
\end{align}
The current-current susceptibility reads
\begin{align}
\chi^{JJ} (0) = \hbar^2 \langle \omega v_1 \vert \omega v_1 \rangle  \; ,
\label{eqn:currentsusceptransportdr}
\end{align}
and the memory function is obtained as
\begin{align}
&\hat{m}^{KK} (i0)  = \frac{i \pi \hbar}{N} \sum_{\vec{k},\vec{k}_1,\vec{k}_2,i} \vert \Psi^{(3)} \left( \! \begin{array}{ccc} i & \zeta & \zeta \\ \vec{k} & \vec{k}_1 & \vec{k}_2 \end{array} \! \right) \vert^2   \notag\\
&\times \sqrt{k_B T n^\prime (\vec{k},i) n^\prime (\vec{k}_1,\zeta) n^\prime (\vec{k}_2,\zeta)}  \notag\\
&\times \big[ \gamma_1 (\vec{k},i) + \gamma_1 (\vec{k}_1,\zeta) + \gamma_1 (\vec{k}_2,\zeta) \big]^2    \notag\\
&\times \big[ \delta (\omega(\vec{k},i) -\omega(\vec{k}_1,\zeta) + \omega(\vec{k}_2,\zeta))  \notag\\
&+ \delta (\omega(\vec{k},i) +\omega(\vec{k}_1,\zeta) - \omega(\vec{k}_2,\zeta))  \notag\\
&+ \delta (\omega(\vec{k},i) - \omega(\vec{k}_1,\zeta) - \omega(\vec{k}_2,\zeta)) \big] \; .
\label{eqn:matrixm2}
\end{align}
Proceeding as outlined above, we find by analytical reasoning that at low $T$ the susceptibility $\chi^{\epsilon \epsilon} (0)$ and $\chi^{JJ}(0)$ vary as $T^2$ and $T^3$ respectively and that $m^{KK}(i0) \propto T^3$. Hence we conclude that the thermal diffusion $\lambda$ vanishes $\propto T$ for $T \to 0$. At high $T$ we obtain $\chi^{\epsilon \epsilon} \propto T$, $\chi^{JJ}(0) \propto T$ and $m^{KK}(i0) \propto T^2$, which means that $\lambda$ vanishes as $T^{-1}$. 

Finally we consider the kinetic coefficients $\lambda^\prime$ and $\pi^\prime$ introduced in Sec. \ref{sec:solutions}. The calculation of $\lambda^\prime$ requires the subtraction of the crystal momentum as a secular variable from the energy current. Eq. (\ref{eqn:lambdacurrenttransport}) has to be replaced by 
\begin{align}
\gamma^\prime_1 (\vec{k}, \alpha) = \hbar [\omega (\vec{k}, \alpha) v_1 (\vec{k}, \alpha) - k_1 g^\lambda ]  \; ,
\label{eqn:currentsusceptransportssr}
\end{align}
where $g^\lambda = \langle k_1 v_1 \vert \omega \rangle / \langle k_1 \vert k_1 \rangle$. The kinetic coefficient $\lambda^\prime$ is then obtained by replacing $\gamma_1$ by $\gamma^\prime_1$ in the expression for $\chi^{JJ}$ and $\hat{m}^{KK}$ that enter Eq. (\ref{eqn:lambdatransportdef}) for $\lambda$. The asymptotic $T$-behavior of $\lambda^\prime$ is the same as that for $\lambda$.

The kinetic coefficient $\pi^\prime_{ij,kl}$ defined in Eq. (\ref{eqn:piprimedef}) plays the role of kinematic viscosity of the phonon gas and the secular variable is the crystal momentum density
\begin{align}
\pi_i (\vec{q})  = \frac{1}{\sqrt{N}} \sum_{\vec{k}, \alpha} \hbar k_i  \: b^{\alpha \dagger}_{\vec{k} - \frac{\vec{q}}{2}} \: b^{\alpha}_{\vec{k} + \frac{\vec{q}}{2}} \; .
\label{eqn:momentumdensitytransport}
\end{align}
Since $\pi^\prime$ is a fourth rank tensor it has the symmetry of the elastic constants. Using again Voigt's notation we may write $\pi^\prime_{pq}$ as:
\begin{align}
\pi^\prime_{pq} = i \chi^{JJ}_{pn} (0) (\hat{m}^{KK} (i0))^{-1}_{nl} \chi^{JJ}_{lq} (0) (\chi^{\pi \pi} (0))^{-1} \; .
\label{eqn:piprimetransportdef}
\end{align}
The crystal momentum susceptibility  reads
\begin{align}
\chi^{\pi \pi} (0) = \hbar^2 \langle k_1 \vert k_1 \rangle = \hbar^2 \langle k_2 \vert k_2 \rangle \; ,
\label{eqn:quasimomentumsusceptibitransport}
\end{align}
The momentum currents are given by
\begin{align}
J_n = \frac{1}{\sqrt{N}} \sum_{\vec{k},\alpha} \gamma_n (\vec{k}, \alpha) b^{\alpha \dagger}_k b^{\alpha }_k  \; ,
\label{eqn:corrientepi}
\end{align}
with
\begin{align}
\gamma_1 (\vec{k}, \alpha) &= \hbar [ v_1 (\vec{k}, \alpha) k_1 - \omega (\vec{k}, \alpha) g^{\pi}_{11} ]  \notag\\ 
\gamma_2 (\vec{k}, \alpha) &= \hbar [ v_2 (\vec{k}, \alpha) k_2 - \omega (\vec{k}, \alpha) g^{\pi}_{22} ] \notag\\ 
\gamma_6 (\vec{k}, \alpha) &= \hbar  v_2 (\vec{k}, \alpha) k_1 = \hbar  v_1 (\vec{k}, \alpha) k_2  \; ,
\label{eqn:gammaspiprime}
\end{align}
where $g^{\pi}_{11} = g^{\pi}_{22} = \langle \omega  \vert k_1 v_1 \rangle / \langle \omega \vert \omega \rangle$. The static current-current susceptibilities are
\begin{align}
\chi^{JJ}_{np} (0) = \langle \gamma_n \vert \gamma_p \rangle  \; ,
\label{eqn:currentcurrentpiprimesus}
\end{align}
and the relevant elements of the memory function are obtained as
\begin{align}
&\hat{m}^{KK}_{nl} (i0)  = \frac{i \pi \hbar}{ N} \sum_{\vec{k},\vec{k}_1,\vec{k}_2,i} \vert \Psi \left( \! \begin{array}{ccc} i & \zeta & \zeta \\ \vec{k} & \vec{k}_1 & \vec{k}_2 \end{array} \! \right) \vert^2  \times \notag\\
&\sqrt{k_B T n^\prime (\vec{k},i) n^\prime (\vec{k}_1,\zeta) n^\prime (\vec{k}_2,\zeta)} \notag\\
&\times \Big\{ \big[ \gamma_n (\vec{k},i) - \gamma_n (\vec{k}_1, \zeta) - \gamma_n (\vec{k}_2, \zeta) \big]   \notag\\
&\times \big[ \gamma_l (\vec{k},i) - \gamma_l (\vec{k}_1,\zeta) - \gamma_l (\vec{k}_2,\zeta)\big] \notag\\ 
&\times \delta (\omega(\vec{k},i) -\omega(\vec{k}_1,\zeta) - \omega(\vec{k}_2,\zeta))  \notag\\
&+ \big[ \gamma_n (\vec{k},i) - \gamma_n (\vec{k}_1, \zeta) + \gamma_n (\vec{k}_2, \zeta) \big]   \notag\\
&\times \big[ \gamma_l (\vec{k},i) - \gamma_l (\vec{k}_1,\zeta) + \gamma_l (\vec{k}_2,\zeta)\big] \notag\\ 
&\times \delta (\omega(\vec{k},i) -\omega(\vec{k}_1,\zeta) + \omega(\vec{k}_2,\zeta))  \notag\\
&+ \big[ \gamma_n (\vec{k},i) + \gamma_n (\vec{k}_1, \zeta) - \gamma_n (\vec{k}_2, \zeta) \big]   \notag\\
&\times \big[ \gamma_l (\vec{k},i) + \gamma_l (\vec{k}_1,\zeta) - \gamma_l (\vec{k}_2,\zeta)\big] \notag\\ 
&\times \delta (\omega(\vec{k},i) +\omega(\vec{k}_1,\zeta) - \omega(\vec{k}_2,\zeta)) \Big\} \; .
\label{eqn:matrixm3}
\end{align}
At low $T$ we find by analytical means that in leading order of $T$, $\chi^{\pi \pi} (0) \propto T$, $\chi^{JJ}(0) \propto T$ and $m^{KK}(i0) \propto T^3$. Hence $\pi^\prime$ diverges as $T^{-2}$ for $T \to 0$. At high $T$ we get $\chi^{\pi \pi} (0) \propto T$, $\chi^{JJ}(0) \propto T$ and $m^{KK}(i0) \propto T^2$, and we conclude that $\pi^\prime$ vanishes as $T^{-1}$. Here and before high $T$ means $T$ larger then $\hbar \omega_D/k_B$ where $\omega_D$ is a BZ boundary frequency. 

Notice that in obtaining the results Eqs. (\ref{eqn:matrixm1}), (\ref{eqn:matrixm2}) and (\ref{eqn:matrixm1}) we do not make use of the concept of a wave-vector dependent relaxation time. The multiple integral expressions are reminiscent of transport theory results obtained by means of variational methods\cite{koler,leibfriedscho,ziman}, however in the present approach we do not resort to any trial function. A numerical evaluation of the transport coefficients is then reduced to a calculation of the single and multiple $\vec{k}$-integrals occurring in the factors $\chi^{AA} (0)$, $\chi^{JJ} (0)$ and $m^{KK} (i0)$. Such a program will be carried  out in the next section where we study the temperature evolution of the transport coefficients.

%
\section{Temperature dependence of transport coefficients}
\label{sec:numresults}
%
As follows from the analytical results of Secs. \ref{sec:solutions} and \ref{sec:transpcoe} the evaluation of the transport coefficients $\eta$, $\lambda$, $\pi^\prime$ and the thermal conductivity $\kappa$ can be achieved by the calculation of the corresponding susceptibilities and  memory functions. For this purpose we need to calculate simple two-folded and triple two-folded integrals inside the BZ. These integrals involve the harmonic frequencies $\omega (\vec{k}, \alpha)$, the group velocities $\vec{v} (\vec{k}, \alpha)$, the anharmonic interactions $\Phi (\vec{k}_1,\alpha_1; \vec{k}_2, \alpha_2; \vec{k}_3, \alpha_3)$, and the momentum and energy conservation functions $\Delta (\vec{k}_1, \vec{k}_2, \vec{k}_3, \vec{G})$ and $\delta (\omega (\vec{k}_1, \alpha_1), \omega (\vec{k}_2, \alpha_2), \omega (\vec{k}_3, \alpha_3) )$. We will use the corresponding input parameters for graphene\cite{seba}.

Following the steps of Ref. (\onlinecite{seba}) we first generate the hexagonal BZ for graphene and then for each of the $N$ $\vec{k}$-points in the BZ we calculate the acoustic phonon dispersions by direct diagonalization of the dynamical matrix\cite{mohr}. Because of band-crossing, we also make use of an auxiliary algorithm to keep the correct sorting of phonon modes after diagonalization. \\

\begin{figure}
\includegraphics[trim = 0.5cm 0cm 0cm 0.0cm, clip, width=8.7cm]{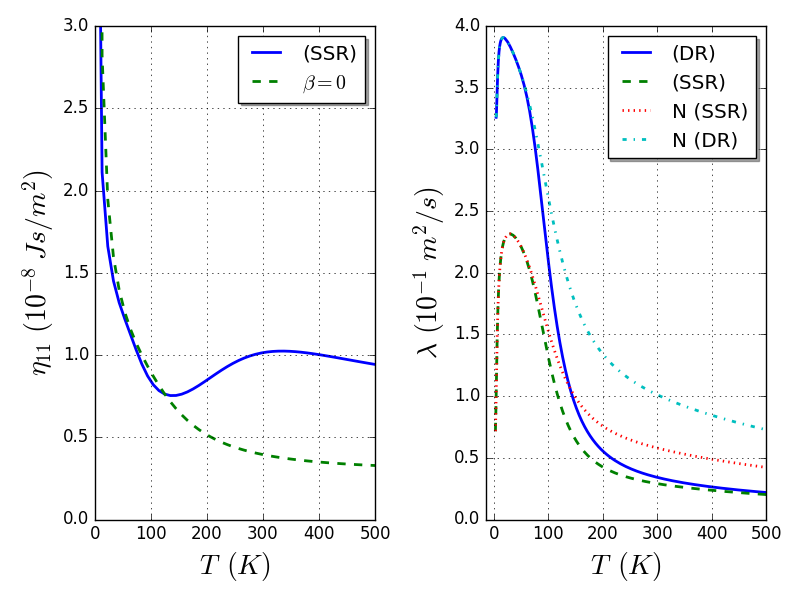}
\caption{Left panel: dynamic viscosity $\eta_{11}$ calculated for graphene according to Eqs. (\ref{eqn:defcl}) and (\ref{eqn:etatransportdef}). As mentioned in the text, the values of $\eta_{11}$ in solid blue line are the same in both DR and SSR. As a comparison, in dash green line we plot the calculations with $\beta = 0$. Right panel: thermal diffusivity $\lambda$ (in the DR) and $\lambda^\prime$ (in the SSR) calculated for graphene according to Eq. (\ref{eqn:lambdatransportdef}) in solid blue and dash green lines respectively. In dot blue and dot green lines we show the corresponding N-processes contributions.}
\label{fig:eta-lambda}
\end{figure}

\begin{figure}
\includegraphics[trim = 0.5cm 0cm 0cm 0.0cm, clip, width=8.7cm]{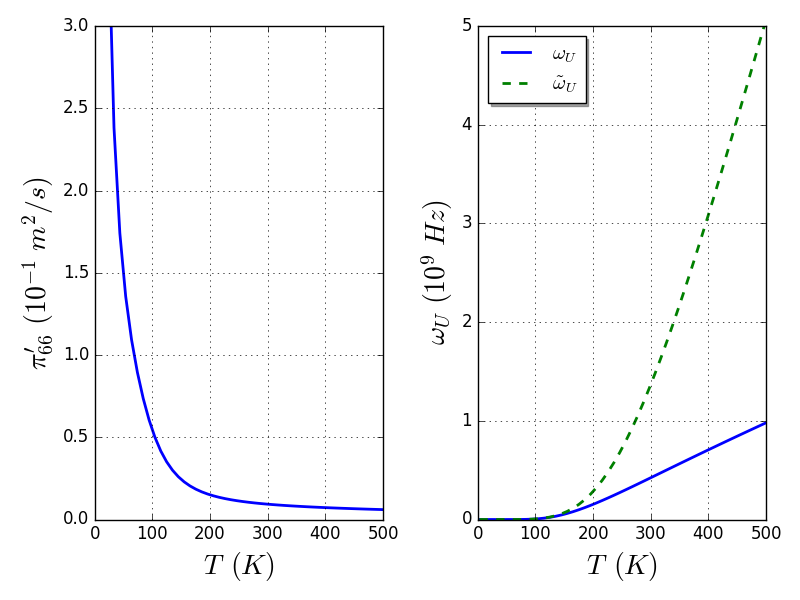}
\caption{Left panel: kinetic phonon viscosity $\pi^\prime_{66}$ calculated according to Eq. (\ref{eqn:piprimetransportdef}). 
Right panel: relaxation frequencies $\omega_U$ and $\tilde{\omega}_U$  calculated for graphene from Eq. (\ref{eqn:omega_U_calc_2}) with our original Hamiltonian in solid blue line and with the inclusion of in-plane scattering in dash green line.}
\label{fig:piprime-omegaU}
\end{figure}

\begin{figure}
\includegraphics[trim = 0.5cm 0cm 0cm 0.0cm, clip, width=8.9cm]{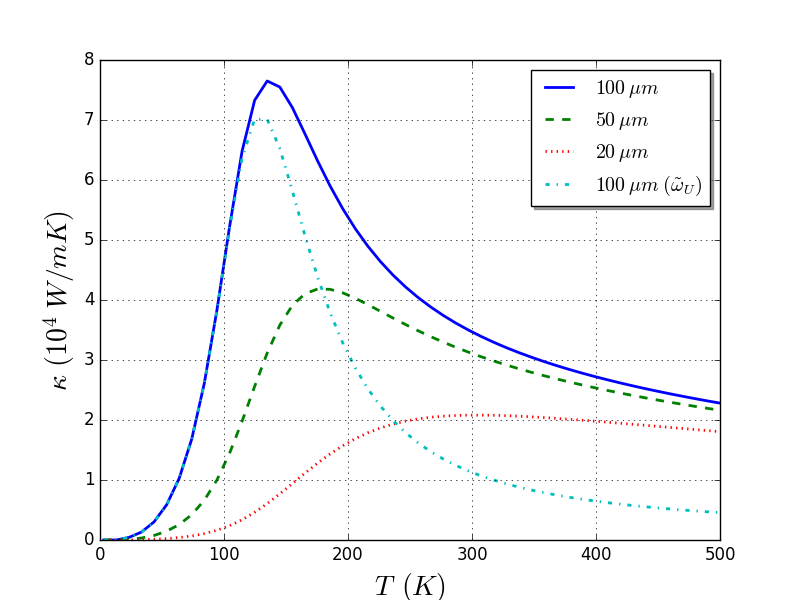}
\caption{Thermal conductivity $\kappa$ calculated for rectangular graphene samples with widths $w= 100$, $50$, and $20 \: \mu$m according to Eq. (\ref{eqn:kappapoiseuille}). In dot-dash cyan line we show the behavior of $\kappa$ with $w=100 \: \mu$m including in-plane scattering in the calculation of $\omega_U$, i.e. by replacing $\omega_U$ with $\tilde{\omega}_U$. }
\label{fig:kappaU}
\end{figure}

The evaluation of simple two-folded integrals for the calculation of scalar products, e.g. $\beta$ and $d$, or susceptibilities, e.g. $\chi^{JJ} (0)$ and $\chi^{\epsilon \epsilon} (0)$, is performed in a straightforward way by summing over all $\vec{k}$-points and the three acoustic polarizations. In Figs. \ref{fig:ssvel}, \ref{fig:eta-lambda}, \ref{fig:piprime-omegaU} and \ref{fig:kappaU} the number of $\vec{k}$-points used is $N=25600$. Triple two-folded integrals are required specifically for memory functions $\hat{m}^{KK} (i0)$. In order to evaluate them we generate a set of all possible combinations $\{ \vec{k}_1, \vec{k}_2, \vec{k}_3 \}$ from the original list of $\vec{k}$-points in the BZ. Among all these combinations, those which satisfy $ \vec{k}_1 + \vec{k}_2 + \vec{k}_3 = 0$ belong to the N-processes group and those which satisfies $ \vec{k}_1 + \vec{k}_2 + \vec{k}_3 = \vec{G}$, for any reciprocal vector $\vec{G}$, belong to the U-processes group. The explicit distinction between normal and umklapp groups allows us to determine their contributions separately, as it will be shown below. Energy conservation is enforced by means of a Gaussian function $(\epsilon \sqrt{\pi})^{-1} \: \textrm{exp} [(-\omega (\vec{k}_1, \alpha_1) + \omega (\vec{k}_2, \alpha_2) + \omega (\vec{k}_3, \alpha_3)) / \epsilon]^2$, where a broadening parameter $\epsilon = 5$ cm$^{-1}$ is chosen. Memory functions are then evaluated by summing over normal and umklapp groups. 

Left panel of Fig. \ref{fig:eta-lambda} shows in solid blue line the temperature evolution of the dynamic viscosity $\eta_{11}$ calculated for graphene according to Eqs. (\ref{eqn:etatransportdef} - \ref{eqn:matrixm1}). Because of the even $\vec{k}$-dependence of the thermoelastic coupling $h_{ij} (\vec{k}, \zeta)$, the scalar products $\langle h_{ij} \vert \chi^k \rangle$ for $k= \{ 1,2\}$ in Eq. (\ref{eqn:viscositytensordef}) are zero. This makes the summation condition  $l > 2$ equivalent to $l > 0$, resulting in $\eta_{ij}$ having the same value in both diffusive and second sound regimes. As observed the dynamic viscosity shows a steep increase with $T \to 0$ in accordance with the $T$-dependence discussion in the previous section and then presents a local minimum at $T \sim 140$ K. The reason of this local minimum is related to the temperature dependence of the thermal tension $\beta$, which remains negative according to Eq. (\ref{eqn:betadef}) (the change in sign of $\beta$ at higher temperatures is a direct consequence of in-plane scattering\cite{seba}). Right panel shows the values of the thermal diffusion coefficients $\lambda$ and $\lambda^\prime$ calculated according to Eqs. (\ref{eqn:energydensitytransport} - \ref{eqn:currentsusceptransportssr}). In addition we plot the corresponding N-processes contributions to $\lambda$ and $\lambda^\prime$. As expected the low $T$ behavior in both the diffusion and second sound regimes is dominated entirely by N-processes. The contribution of U-processes becomes only noticeable above $T \sim 60$ K. \\

The kinematic viscosity coefficient $\pi^\prime_{66}$ for graphene, that enters in the expression for the thermal conductivity $\kappa$ Eq. (\ref{eqn:kappapoiseuille}), is plotted in the left panel of Fig. \ref{fig:piprime-omegaU}. According to Eqs. (\ref{eqn:momentumdensitytransport} - \ref{eqn:matrixm3}), this coefficient presents a steep increase with $T \to 0$ and a monotonically decreasing behavior for increasing $T$. Right panel of Fig. \ref{fig:piprime-omegaU} shows the temperature evolution of the relaxation frequency $\omega_U$ calculated according to Eq. (\ref{eqn:omega_U_calc_2}) and the values of $\tilde{\omega}_U$, i.e. the relaxation frequency $\omega_U$ with the addition of the in-plane scattering terms Eqs. (\ref{eqn:defomegainplane} - \ref{eqn:defpsimayusinplane}). At low $T$ both frequencies are essentially the same, meaning that in-plane scattering can be completely neglected. With $T$ increasing above $\sim 150$ K the relative difference between $\omega_U$ and $\tilde{\omega}_U$ becomes noticeable and in particular it reaches a value of $ \sim 70 \%$ at $T = 300$ K.

Fig. \ref{fig:kappaU} shows the thermal conductivity $\kappa$ calculated for graphene according to Eq. (\ref{eqn:kappapoiseuille}) for three different widths $w = 100$, $50$, and $20$ $\mu$m. At low $T$, $\kappa$ behaves as predicted by Eq. (\ref{eqn:kappapoiseuille2}) and tends to zero with $\pi^\prime_{66} \to \infty$. With increasing $T$, $\pi^\prime_{66}$ decreases and, as far as the relaxation frequency $\omega_V$ of crystal momentum destroying processes is small, $\kappa$ increases. At high $T$ where U-processes are dominant (we assume that impurity scattering is negligible and hence $\omega_V \equiv \omega_U$), $\kappa$ decreases with increasing $T$, in accordance with Eq. (\ref{eqn:kappapoiseuille3}). In the intermediate temperature regime $\kappa$ reaches a maximum that is determined by the subtle interplay of the $T$-dependence of $\pi^\prime_{66}$ and $\omega_U$. The maximum shifts to higher $T$ with lower sample width $w$, in agreement with Refs. \onlinecite{fugallo,xunhydro}.

A quantitative comparison with the results of other works\cite{cepe_naturecom,lee_naturecom,fugallo,xunhydro} shows that our values of $\kappa$ for $T$ above $150$ K are larger. Also the inclusion of in-plane scattering does not resolve this discrepancy. A further reduction of the values of $\kappa$ requires the inclusion of impurity scattering and possibly of fourth order scattering processes of flexural (ZA) phonons\cite{fourthdegradation}. We expect that the latter lead to a reduction of $\pi^\prime_{66}$ or equivalently to a decrease of the mean free path $l_N$ for N-processes.
%

%
\section{Conclusions}
\label{sec:conclusions}
%
Starting from coupled dynamic equations for in-plane lattice displacements (\ref{eqn:soundwaves}) and flexural phonon kinetics (\ref{eqn:soundwaves}) we have used perturbative methods to obtain the hydrodynamic equations describing elastic sound waves and local temperature fluctuations. We have distinguished two scenarios: (i) if phonon energy is the only conserved quantity, temperature fluctuations are described by a diffusion equation; (ii) if in addition crystal momentum is taken as an almost conserved quantity and the frequency window condition is fulfilled\cite{prohofsky}, temperature fluctuations are described by a wave equation, the so-called second sound. In the zero frequency limit and in presence of a constant temperature gradient, scenario (ii) leads to Poiseuille flow.

The present study differs from previous theoretical works on the thermal properties of 2D crystals (see references in Sec. \ref{sec:introduction}) in several aspects. We start from coupled dynamic equations derived from a microscopic Hamiltonian\cite{SMP} and obtain therefrom a unified description for elastic and thermal hydrodynamic phenomena. The coupling is mediated by the elastic tension coefficient which is an anharmonic effect due to the coupling between in-plane and flexural lattice displacements. By investigating the elastic and thermal response functions, we have shown how the thermal resonances (second sound or Landau-Placzek peak) appear in the displacement-displacement response function while conversely the sound wave doublet is present in the thermal response. The interplay of elastic and thermal phenomena in various scattering laws is a challenge for new experiments. Indeed most recently second sound has been observed in graphite at temperatures above 100 K by means of time-resolved optical measurements\cite{ssgraphite}. The method is based on the transient thermal grating technique\cite{ttgrating}. The spatial and temporal decay of the thermal grating by second sound heat transport is reflected, as a consequence of thermo-mechanical coupling, in the surface displacement field that acts as a transient diffraction grating.

A further distinct aspect of our work is the calculation of various transport coefficients such as the first sound viscosity $\eta$, the thermal diffusion $\lambda$ and the phonon viscosity $\pi^\prime$. Together with the crystal momentum relaxation frequency $\omega_V$ these coefficients characterize the broadening of elastic and thermal resonances and determine the strength of the thermal conductivity $\kappa$. Using Kubo-Mori response theory we have written the transport coefficients as relaxation functions of generalized currents of secular variables and calculated the current-current relaxation functions by means of an equation of motion method due to G\"otze and one of the present authors\cite{GM3}. Thereby we have obtained closed expressions for the transport coefficients in form of multiple integrals over the Brillouin zone that are reminiscent of variational theory results \cite{leibfriedscho,ziman,sevic}. A quantitative evaluation of the transport coefficients and their $T$-dependence has been performed by numerical calculations in Sec. \ref{sec:numresults}. In addition we have study the asymptotic behavior at low and high temperatures by analytical means. In particular we find that the viscosities $\eta$ and $\pi^\prime$ diverge at low $T$ while the thermal diffusion $\lambda$ tends to zero. 
Special care has been devoted to the calculation of Poiseuille flow\cite{gurzhi} and the static thermal conductivity $\kappa$. We have solved the differential equation (\ref{eqn:poisoille1}) that describes phonon drift in presence of a static temperature gradient for a rectangular 2D crystal with width w. The thermal conductivity, Eq. (\ref{eqn:kappapoiseuille}), then depends on $\omega_V$ and $\pi^\prime_{66}$. At low $T$ and for a sample with finite width, it follows from Eq. (\ref{eqn:kappapoiseuille2}) and from the already mentioned divergence of $\pi^\prime$ that $\kappa$ vanishes with $T \to 0$. At high $T$ where $\omega_V$ is large and $\pi^\prime_{66}$ vanishes, Eq. (\ref{eqn:kappapoiseuille3}) shows that $\kappa$ tends asymptotically to zero with increasing $T$. The complete temperature evolution of $\kappa$ is presented in Fig. (\ref{fig:kappaU}). We recall that $\pi^\prime_{66}$ is due to N-processes, see Eq. (\ref{eqn:piprimedef}) and remark following Eq. (\ref{eqn:ajpoiseuille}). The vanishing of $\kappa$ is then the ultimate consequence of the disappearance of thermal phonons as heat carriers in the zero temperature limit.

%
\section{Acknowledgements}
\label{sec:Aknowledgements}
%

This work was supported by the Flemish Science Foundation (FWO-Vl).

\bibliographystyle{unsrt}
\bibliography{bibi}
%

\end{document}